%% file: EGauthorGuidelines-cgf-sub.tex
\documentclass{egpubl}
 
\JournalSubmission    % uncomment for submission to Computer Graphics Forum
\usepackage[T1]{fontenc}
\usepackage{dfadobe}  

\usepackage{cite}  % comment out for biblatex with backend=biber
\BibtexOrBiblatex
\electronicVersion
\PrintedOrElectronic
\ifpdf \usepackage[pdftex]{graphicx} \pdfcompresslevel=9
\else \usepackage[dvips]{graphicx} \fi

\usepackage{egweblnk} 
\usepackage{amsmath,amsfonts}
\usepackage{algorithmic}
\usepackage{algorithm}
\usepackage{array}
\usepackage{textcomp}
\usepackage{stfloats}
\usepackage{url}
\usepackage{verbatim}
\usepackage{cite}
\usepackage{caption}
\usepackage{subcaption}
\usepackage{graphicx}
\usepackage{multirow}
\usepackage{array}
\usepackage{colortbl}
\usepackage{xcolor}
\usepackage{makecell} % in preamble
\usepackage{soul}
\usepackage{booktabs}
\title[Incidental Visualizations: Augmented Reality as a Medium for Contextual Information]%
      {Incidental Visualizations:\\ Augmented Reality as a Medium for Contextual Information}

\author[M. Heitor, J. Moreira, \& D. Gonçalves]
{\parbox{\textwidth}{\centering
        M. Heitor$^{1}$,
        J. Moreira$^{1,2}$\orcid{0000-0002-5461-5217},
        and D. Gonçalves$^{1,2}$\orcid{0000-0002-5121-6296}
         }
        \\
{\parbox{\textwidth}{\centering $^1$Instituto Superior Técnico, University of Lisbon, Av. Rovisco Pais 1, 1049-001 Lisboa, Portugal\\
        $^2$INESC-ID, R. Alves Redol 9, 1000-029 Lisboa, Portugal
       }
}
}

\begin{document}

% uncomment for using teaser
% \teaser{
%  \includegraphics[width=0.9\linewidth]{eg_new}
%  \centering
%   \caption{New EG Logo}
% \label{fig:teaser}
%}

\maketitle
%-------------------------------------------------------------------------
\begin{abstract}
   In today’s fast-paced world, delivering information efficiently and unobtrusively is essential. While ambient and glanceable visualizations provide real-time data, they can increase cognitive load and disrupt primary tasks. We investigate incidental visualizations (IVs), a novel concept in information visualization designed to present contextually relevant information briefly and spontaneously, with minimal user interaction. Augmented Reality (AR) offers an ideal medium for this integration, embedding visualizations directly within the user’s environment. Through controlled user studies on logic-based game tasks (Sudoku and Connect 4), this work compares ambient, periodic, and incidental visualization patterns in terms of comprehension accuracy, performance, and disruption. Results indicate that IVs deliver information as effectively as ambient displays while minimizing disruption, highlighting their potential for adaptive, context-aware information delivery in AR environments.
   
\begin{CCSXML}
<ccs2012>
   <concept>
       <concept_id>10003120.10003121.10003124.10010392</concept_id>
       <concept_desc>Human-centered computing~Mixed / augmented reality</concept_desc>
       <concept_significance>500</concept_significance>
       </concept>
   <concept>
       <concept_id>10003120.10003121.10003122.10003334</concept_id>
       <concept_desc>Human-centered computing~User studies</concept_desc>
       <concept_significance>500</concept_significance>
       </concept>
   <concept>
       <concept_id>10003120.10003145.10011769</concept_id>
       <concept_desc>Human-centered computing~Empirical studies in visualization</concept_desc>
       <concept_significance>500</concept_significance>
       </concept>
 </ccs2012>
\end{CCSXML}

\ccsdesc[500]{Human-centered computing~Mixed / augmented reality}
\ccsdesc[500]{Human-centered computing~User studies}
\ccsdesc[500]{Human-centered computing~Empirical studies in visualization}

\printccsdesc   
\end{abstract}  
%-------------------------------------------------------------------------

\input{sections/introduction}
\input{sections/related_work}
\input{sections/methodology}
\input{sections/results}
\input{sections/conclusions}

\section*{Acknowledgments}
This work was developed within the scope of the UNESCO Chair on AI \& VR, and supported by national funds through Fundação para a Ciência e a Tecnologia, I.P. (FCT) under projects UID/50021/2025 and UID/PRR/50021/2025. The authors used AI for language editing. All content was reviewed and verified by the authors, who take full responsibility for the final manuscript.

%-------------------------------------------------------------------------
% bibtex
\bibliographystyle{eg-alpha-doi} 
\bibliography{egbibsample}       

\end{document}

%% file: sections/introduction.tex
\section{Introduction}
In our fast-paced world, information is everywhere, constantly surrounding us both physically and digitally. Therefore, we need accessible ways to interpret it without becoming overwhelmed.
Pioneered by Mark Weiser in the 1990s \cite{Weiser:1993}, the movement of Ubiquitous Computing envisioned technology fading into the background, blending seamlessly into everyday life. The philosophy of Calm Technology \cite{Weiser_Brown:1996} suggests computing would be effective, invisible, and informative without seizing our focus.
Following this ideal to fully integrate these high amounts of information, we need a way to convey data without demanding the user's full attention. However, the current methods differ in their approach and purpose.

Different approaches have emerged, namely ambient visualizations \cite{Ishii:1998}, which persistently exist embedded within the environment (e.g., airport dashboards). Another approach is glanceable visualizations \cite{Blascheck:2021.2_Characterizing_GlanceableVis}, crafted to be perceived within brief exposure times (e.g., smartwatch charts). 

The embedded nature of these approaches, along with their need for deep spatial integration, highlights AR as a promising medium for these visualizations \cite{ Lu:2021.1_EvaluatingPotential, Daskalogrigorakis:2021, Davari:2020, Piening:2021, Iquiapaza:2023}. AR integrates virtual placement of information within the environment, with no extra physical hardware for each visualization.

Under a recent surge of development, previous processing limitations \cite{Lu:2021.1_EvaluatingPotential, Pohl:2024, KleinSedlmairSchreiber:2022} have been surpassed, making AR increasingly viable for everyday use \cite{Lu:2021.1_EvaluatingPotential, Davari:2022}.

However, this now also allows for the ability to show unrestricted amounts of information, which can quickly lead to occlusion and visual clutter \cite{Davari:2020}. 

Despite their convenience, glanceable and ambient visualizations can introduce challenges by diverting attention from primary tasks. They often require users to shift their focus to receive the information, acting as an on-demand mechanism and interrupting ongoing activities to perceive or interact with the visualizations. This can be problematic in situations such as driving or operating heavy machinery, where users cannot afford to spend too much time analyzing information or searching for the necessary details. 

IVs \cite{Moreira:2020} offer an alternative to the previous settings, appearing automatically during a single glance and only when deemed relevant for the task at hand, without requiring any user input or request \cite{Moreira:2023a,Moreira:2023b}. They appear during main tasks, as users expect to get relevant information when needed, with minimal disruption.
IVs uniquely combine brief, contextually situated visualizations that appear directly in the user’s field of view, minimizing occlusion, avoiding interaction, and reducing disruption to the primary task. This balance of contextual relevance and unobtrusiveness positions IVs as candidates for AR applications.

While Moreira et al.'s founding work on IVs \cite{Moreira:2020, Moreira:2023a, Moreira:2023b, Moreira:2024, Moreira:2024b_IVvsAMB} sets initial design guidelines and establishes a strong foundation, a comprehensive design framework remains absent, potentially influenced by the current lack of defined everyday applications. Moreover, this field is still in its early stages, requiring further research to establish clear implementation guidelines and use-case scenarios to fully realize their potential utility.

This work contributes to the development of incidental visualizations by further testing and extending the design guidelines proposed by Moreira et al. Through a controlled AR-based user study over distinct tasks (Sudoku and Connect Four), it evaluates how IVs compare to ambient, periodic, and no visualization conditions in terms of comprehension, disruption, and task performance.
The study provides evidence on whether brief, contextually triggered cues can serve as a viable alternative to other visualization philosophies, supporting informativeness while minimizing attention load and visual clutter. Additionally, by integrating task-specific visualization design and world-fixed AR placement, the work offers insights that inform the development of a broader design framework for IVs and their applicability across diverse real-world scenarios.

%% file: sections/related_work.tex
\section{Related Work}

AR overlays digital information onto the physical world, enhancing real-world experiences \cite{Martins:2022} and preserving links between data and physical referents, forming the basis for Situated Visualizations (SVs) \cite{KleinSedlmairSchreiber:2022}. SVs integrate data with the environment and physical referents \cite{Willett:2017}, providing contextually relevant insights to support understanding, allowing users to focus on primary tasks \cite{Martins:2022, Kraus:2021}. Their conceptual model connects source to representation, encompassing spatial, temporal, and social situatedness \cite{Bressa:2021, Kraus:2021}. Among visualization types, panels are the most common AR visualization pattern, with mirror and morph variants used less frequently \cite{Lee:2024}.

Empirical studies \cite{Lu:2021.1_EvaluatingPotential, Jeffri:2021} show that SVs in AR reduce cognitive load and enhance real-time decision-making by removing the need to map between digital and physical referents \cite{Martins:2022, Marques:2019, Bressa:2021}. Yet, despite challenges like occlusion and overload still persisting, which limit their effectiveness \cite{Kraus:2021, Ens:2021}, SVs remain a promising foundation for immersive analytics in AR, enabling users to interpret and act on information directly within their physical context.

\subsection{Ambient Visualizations}
The philosophy of Calm Technology \cite{Weiser_Brown:1996} envisions technologies that shift information between the periphery and the center of attention. Rooted in Weiser’s earlier concept of ubiquitous computing \cite{Weiser:1993}, the idea is that pervasive computing systems should enhance, not interrupt, human experience.
Derived from these ideas, peripheral visualizations communicate information with minimal disruption, evolving into ambient visualizations that embed data into the physical environment using subtle cues such as light, color, or motion \cite{Ishii:1998, Moreira:2023a}. These visualizations are continuously available and aim to provide information naturally, without requiring focused attention, through subtle awareness \cite{Blascheck:2021.2_Characterizing_GlanceableVis}.

Therefore, ambient visualizations excel when continuous monitoring is necessary, allowing users to remain aware while engaged in other activities.
They are typically non-intrusive and abstract, with aesthetics often serving as a key design feature \cite{Bressa:2021}.
Despite their non-intrusive intent, they can still contribute to occlusion and visual overload, potentially distracting users from primary tasks. As they focus on persistent availability rather than context-driven temporal visualizations, cognitive load presents as an emerging issue. This availability of information introduces the potential of layering unrestricted digital information from multiple visualization sources, increasing cognitive load \cite{Moreira:2023a}. 

\subsection{Glanceable Visualizations}
While ambient visualizations focus on persistent availability, as technology shifted toward mobile and wearable devices, glanceable visualizations emerged with the aim of reducing obtrusiveness \cite{Lu:2021.1_EvaluatingPotential}, prioritizing the focus on the real world where information can be quickly accessed through a glance at these devices.
They allow users to retrieve relevant data at a glance with little focus break from the physical world. In AR, glanceable visualizations overlay content directly within the user’s view, enabling efficient interactions that enhance everyday task performance \cite{Davari:2022, Lu:2021.2_Exploration_Techniques}.

Glanceable AR systems require lightweight, rapid activation methods that avoid visual clutter and occlusion \cite{Davari:2020, Lu:2021.2_Exploration_Techniques} and techniques such as visual tokens \cite{Piening:2021} and automatic-translucency policies \cite{Davari:2020} which prioritize real-world visibility. Interaction mechanisms range from gaze triggers to hand gestures, though challenges such as the Midas Touch Problem and user fatigue remain \cite{Lu:2021.2_Exploration_Techniques, Iquiapaza:2023}.
To mitigate occlusion and cognitive load, some systems exploit temporal situatedness, brief visualizations of information that can be perceived pre-attentively \cite{Lu:2021.1_EvaluatingPotential, Moreira:2020}. These low-interference visualizations convey data efficiently, aligning with the goals of IV in AR, to support perception without interrupting primary tasks.

\subsection{Incidental Visualizations}

Incidental visualizations (IVs), introduced by Moreira \cite{Moreira:2023b}, deliver timely, context-triggered information that supports users’ ongoing tasks without requiring interaction or diverting attention. Unlike glanceable visualizations, seen on smartwatches or AR systems \cite{Lu:2021.1_EvaluatingPotential, Moreira:2023a}, which demand explicit user action, IVs appear automatically based on contextual cues, such as environment or task state \cite{Davari:2022, Ens:2021}.

IVs combine principles from situated, ambient, and glanceable visualizations, emphasizing temporal relevance, minimal cognitive load, and seamless real-world integration \cite{Marques:2019, Moreira:2024}. Moreira et al.'s founding work on IVs sets initial design guidelines, proving their effectiveness and usability during primary tasks \cite{Moreira:2023a, Moreira:2024}, setting perception range limitations \cite{Moreira:2020}, proposing placement and presentation specifications \cite{Moreira:2023b} and comparing their effectiveness next to the established ambient visualizations \cite{Moreira:2024b_IVvsAMB}. These studies strongly suggested that users can perceive short, low-complexity data bursts effectively \cite{Moreira:2020} and maintain task performance while gaining incidental awareness \cite{Moreira:2023a, Moreira:2024}. Comparative work confirmed IVs’ suitability for focus-intensive activities where continuous visualizations would cause distraction or clutter \cite{Moreira:2024b_IVvsAMB}.

Through automated, context-driven activation and short exposure times, IVs minimize occlusion and interaction demands, positioning them as a natural evolution of glanceable visualizations and a promising approach for AR applications.

\subsection{Discussion}

The integration of AR into everyday environments \cite{Martins:2022, Lu:2021.2_Exploration_Techniques} offers new opportunities to fit tasks that require seamless information delivery. Existing paradigms, such as ambient and glanceable visualizations, provide peripheral awareness and rapid data access but are prone to cognitive overload, occlusion, and demands on attention or interaction \cite{Guarese:2020, Blascheck:2021.1, Davari:2020}. IVs differ by appearing contextually and briefly, requiring no interaction and causing minimal cognitive disruption \cite{Moreira:2020, Moreira:2023a, Moreira:2023b}, making them well-suited to specific scenarios where the lack of these characteristics is detrimental, and particularly for AR. However, empirical research on IVs remains scarce, with open questions regarding timing, placement, and real-world applicability \cite{Moreira:2024, Moreira:2024b_IVvsAMB}. This study investigates IVs in AR, aiming to explore AR as a potential tool to support this information delivery and settings that balance informativeness with minimal disruption, ensuring they enhance user performance without adding cognitive burden.

%% file: sections/methodology.tex
\section{Methodology}
The objective of this work was to evaluate the effectiveness of IVs in supporting real-world tasks within AR. Two logic-based games, Sudoku and Connect Four, were used in a within-subjects study comparing IVs and validating their effectiveness against other differing patterns, which represent distinct philosophies of attention. Each scenario measured comprehension accuracy, task performance, and disruption levels, with the goal of testing design guidelines for future AR applications of IVs.

We compared IVs against ambient visualizations, an established static pattern recognized for reliably retrieving information through persistent visualizations, despite being prone to disruption challenges inherent in its continuous, embedded presence. To assess the impact of timeliness in IV, we included a periodic display condition, where the visualizations are triggered at regularly set intervals, ignoring context relevance. Finally, we also considered a None condition, which presented no visualizations, as a control condition to allow for the test of task suitability and information gain as well as a baseline.

\subsection{Research Questions and Hypothesis}
Our research questions and hypotheses fell under the scope of three different categories: comprehension accuracy, primary task disruption, and task performance. A summary is presented in Table \ref{tab:research_questions}.

\subsubsection{Comprehension Accuracy}
Comprehension accuracy, a central metric in InfoVis, measures how effectively users extract and interpret data from visual representations \cite{Moreira:2020, Cleveland:1984}. It reflects both perceptual clarity and cognitive efficiency, often studied alongside response time and cognitive load to evaluate visualization performance. In time-constrained or attention-limited scenarios, accuracy depends on how well information is conveyed within short exposure times and minimal user interaction \cite{Blascheck:2021.1, Moreira:2023a}. This motivated our first research question.

\textbf{RQ1:} \textit{How do different visualization types (none, ambient, periodic, and incidental) influence comprehension accuracy?}

This metric was evaluated through the acquisition of information that was implicitly generated by the main task. 
This derived information regarding the current game state was explicitly represented in the visualizations, enabling immediate assessment of data comprehension, regardless of the user's performance in the primary activity. The implicit nature meant that although the main task was the focus, the information needed for the accuracy metric was a complementary insight derived from the main task's progression.

Timed visualization approaches, including incidental visualizations, have shown promise for delivering relevant cues without disrupting primary tasks \cite{Carswell:1987, Moreira:2023b}. Therefore, accuracy differences can be explored regarding contextually timed visualizations, when deemed relevant for the task, which are expected to contribute to better information acquisition compared to visualizations with non-dynamic preset timings and a periodic pattern. In contrast, always-on ambient visualizations may increase distraction or cognitive overload due to their continuous presence \cite{Davari:2022}. We expected the effectiveness of having visualizations in relation to the accuracy task to perform better than no visualization, as any kind of visualization yields more accurate results than the lack thereof. Integrating these visualization types into AR allowed for context-sensitive and adaptive information delivery, potentially mitigating occlusion and supporting comprehension in dynamic environments \cite{Piening:2021, Lu:2021.1_EvaluatingPotential}. These considerations supported the definition of the following hypotheses.

\textbf{H1:} \textit{Ambient, incidental, and periodic visualizations improve comprehension accuracy compared to no visualizations.}

\textbf{H2:} \textit{Incidental visualizations provide higher comprehension accuracy than periodic visualizations by delivering information at contextually relevant moments.}

\textbf{H3:} \textit{Incidental visualizations provide  comprehension accuracy close to that of ambient visualizations, despite being temporarily situated.}

\subsubsection{Primary Task Disruption}

Disruption testing examines how visualizations affect user attention, task performance, and cognitive load \cite{Moreira:2020}. In AR, where virtual information overlays the physical environment, visualizations may compete with real-world stimuli, amplifying attention demands \cite{Piening:2021}. To assess disruption, participants provided subjective ratings using the unweighted NASA-TLX \cite{Hart:1988} and a Likert-scale question measuring perceived workload and subjective disruption. These qualitative measures captured how secondary visualizations influenced focus on the primary task and comprehension results. This motivated our second question.

\textbf{RQ2:} \textit{How do different visualization types (none, ambient, periodic, and incidental) affect disruption levels during decision-making tasks?}

Prior findings suggest that the absence of visualizations minimizes distraction, while well-designed IVs can integrate seamlessly into user workflows with minimal interference \cite{Moreira:2023a, Moreira:2024}. Periodic and ambient visualizations, though informative, may increase cognitive demand through repetition or constant visibility \cite{Davari:2022}. Integrating these visualization types into AR provided insight into balancing awareness and focus. These considerations led to the following hypotheses.

\textbf{H4:} \textit{Displaying no visualizations results in the lowest disruption levels, compared to the presence of visualizations (ambient, periodic, incidental).}

\textbf{H5:} \textit{Incidental visualizations reduce disruption compared to periodic and ambient visualizations by minimizing time looking away from the primary task.}

\subsubsection{Task Performance}
Task performance reflects how effectively users complete their primary activity while supported by additional but non-crucial visualizations \cite{Moreira:2023a}. The goal of IVs is to assist without degrading performance, offering relevant cues that enhance decision-making or situational awareness when needed. In AR, these cues can improve efficiency by embedding complementary information directly in context \cite{Piening:2021, Moreira:2023a}. This led to our third research question.

\textbf{RQ3:} \textit{How do different visualization types (none, ambient, periodic, and incidental) affect task performance?}

Performance measures were only compared within each participant to account for individual skill differences, isolating visualization effects. We expected visualization-assisted conditions to perform on par with, or better than, the baseline (no visualization), provided perception and usability of the additional information. Ambient visualizations could offer limited benefit due to static or repetitive content, while incidental ones were hypothesized to outperform by delivering timely, relevant information aligned with task context \cite{Moreira:2023a, Moreira:2024, Davari:2022}. These considerations led to the following hypotheses.

\textbf{H6:} \textit{Ambient, incidental, and periodic visualizations maintain primary task performance on par with no visualization.}

\textbf{H7:} \textit{Incidental visualizations outperform other patterns in task performance by supporting with contextually relevant information without unnecessary distraction.}

\begin{table}[t]
\centering
\caption{Summary of Research Questions and Hypotheses}
\label{tab:research_questions}
\setlength{\tabcolsep}{4pt}
\renewcommand{\arraystretch}{1.1}
\begin{tabular}{@{}p{0.7cm}p{0.7cm}p{6.0cm}@{}}
\toprule
\textbf{RQ} & \textbf{H} & \textbf{Description} \\ 
\midrule
RQ1 & H1 & Ambient, incidental, periodic $>$ none visualizations. \\
{}  & H2 & Incidental $>$ periodic, due to contextual relevance. \\
{}  & H3 & Incidental $\approx$ ambient, despite temporary. \\[3pt]
RQ2 & H4 & No visualization $<$ incidental, periodic, ambient. \\
{}  & H5 & Incidental $<$ periodic, ambient in time diversion. \\[3pt]
RQ3 & H6 & Incidental, ambient, periodic $\approx$ none, not disrupting primary task. \\
{}  & H7 & Incidental $>$ ambient, periodic, none, by supporting relevant decisions. \\
\bottomrule
\end{tabular}
\end{table}

\subsection{Primary Tasks}
IVs work on the assumption that the user is familiar with these visual cues and understands that they can provide additional information at any moment, not disrupting the primary task \cite{Moreira:2024}. This means that we assume such visualizations would already be part of people's natural flow in everyday activities. Therefore, to simulate this real-world interaction, we needed a main task that felt intuitive, something engaging yet familiar. Games are commonly used in AR research \cite{KleinSedlmairSchreiber:2022, Daskalogrigorakis:2021, Pohl:2024} to simulate everyday activities. They support both virtual and physical interaction. Additionally, game tasks provide measurable outcomes, giving both ecological validity and control. 

To evaluate IVs under familiar and low-learning conditions while covering both individual and shared settings, two common logic-based games were selected: Sudoku (single-player) and Connect Four (multiplayer). Connect Four is a dynamic, dual-player environment where attention shifts between players and moves and is closer to real-world, collaborative settings. Although it is a multiplayer environment, a computer agent was used as the opponent strategy that the moderator followed in the Connect Four task to ensure scientific control, consistency, and a balanced level of challenge for the participants. Matilde acted as a physical proxy for the agent. In contrast, Sudoku is a single-player-focused activity where attention stays on one area. 

This combination allowed controlled testing across different attention demands and social contexts, aligning with prior studies on glanceable and context-aware AR visualizations \cite{Davari:2022, Iquiapaza:2023}. Sudoku involves filling a 9×9 grid so that each row, column, and sub-grid contains digits 1–9, emphasizing focused, individual reasoning. Connect Four requires two players to alternately drop colored discs in a vertical grid to connect four in a row, introducing dynamic attention shifts and co-presence effects.

Both are simple yet popular games that provided data that could be transformed into supplementary information about the game’s evolving state.
This additional information was displayed as visualizations that dynamically portrayed the current state of the game, monitoring specific metrics in real time. 
For the Sudoku task, the additional game information corresponded to the percentage of correctly filled cells. Then, for the Connect Four task, the information corresponded to the number of possibilities a player had to create a sequence of one, two, three, or four pieces in a row.

\subsection{Information Visualization Task}
In addition to gameplay, a secondary perception task was integrated to measure users’ accuracy in interpreting the visualizations within the AR environment. This assessed perceptual comprehension of supplementary information during both games, ensuring the evaluation of information extraction alongside task performance.

The bar chart visualization had its origin in Moreira’s prior experimental setup \cite{Moreira:2023a, Moreira:2024}, where each bar represented how many sets of connected pieces of length one (single pieces) up to four (sequences that are four-piece long) can be formed with all the player's possible next moves. This supported rapid value comparison through length encoding, a highly accurate perceptual channel \cite{Cleveland:1984}.

For the Sudoku task, while the encoded values were also quantized into four categories, a bar chart was less suited. The data represented the percentage of correct cell entries within each grid quadrant. Therefore, a heatmap directly mapped data to the Sudoku grid, reducing cognitive mapping effort and facilitating spatial reasoning. 
Although prior work showed that green hues provide higher perceptual discrimination \cite{Vanderplas:2020}, AR visualizations can limit color fidelity. 
Therefore, the percentage was represented through linear greyscale, chosen for robustness under varying AR lighting conditions \cite{Vanderplas:2020}. 

\begin{comment}
\begin{figure}[t]
  \centering
  \begin{subfigure}[b]{0.23\textwidth}
    \includegraphics[width=\textwidth]{Images/visConnect4Barchart.png}
    \caption{Barchart visualization.}
    \label{fig:exBarchart}
  \end{subfigure}
  \hfill
  \begin{subfigure}[b]{.23\textwidth}
    \includegraphics[width=\textwidth]{Images/visSudokuHeatmap.png}
    \caption{Heatmap visualization.}
    \label{fig:exHeatmap}
  \end{subfigure}
  \caption{Examples of the visualizations used in the study: the barchart (left) and the heatmap (right).}
  \label{fig:visualizations}
  \vspace{-1em}
\end{figure} 
\end{comment}

Both visualizations employed pre-attentive processing and subitizing principles \cite{Moreira:2023a, Moreira:2024}, limiting elements to four, staying within the subitizing range, to enable rapid, unconscious perception of extrema (e.g., highest bar or darkest cell). 
However, the ease of identifying each maximum could be influenced by the perceptual difference between the highest and second-highest values. When these values are close, asking which is higher, for example, could become ambiguous, approaching the threshold of the just-noticeable difference. To account for this, an ambiguity factor was introduced, quantifying the relative difference between values and providing an objective measure of task difficulty independent of participant performance. This metric not only explains variations within highly accurate channels, such as bar length, but is also critical when analyzing weaker encodings, such as color luminance for the heatmap visualizations, where performance is inherently less precise. More details on this factor in the next section.

\subsection{User Study}
Each game context also supported a secondary task to evaluate the perception of additional information related to the game. In both game contexts, this task consisted of an extremum identification task (“Find Extremum”), selecting the maximum value in each visualization. This allowed us to validate if people were able to understand the information conveyed via the several visualization modes.

\begin{figure}[t]
    \centering
    \includegraphics[width=\linewidth]{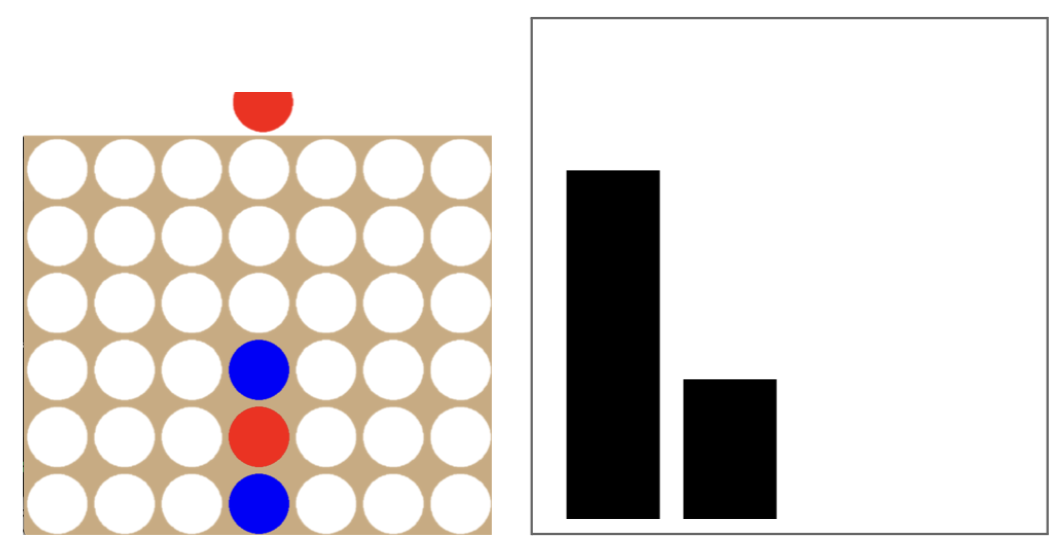}
    \caption{Example of a Connect Four board and the corresponding barchart visualization.}
    \label{fig:exConnect4Vis}
    \vspace{-1em}
\end{figure}

For Connect Four, participants were required to identify the connected piece length for which more combinations were possible in the player's next move. All valid plays were computed per board state, and the tallest bar in the bar chart encoded the corresponding maximum value (Fig.~\ref{fig:exConnect4Vis}).

\begin{figure}[b]
    \centering
    \includegraphics[width=\linewidth]{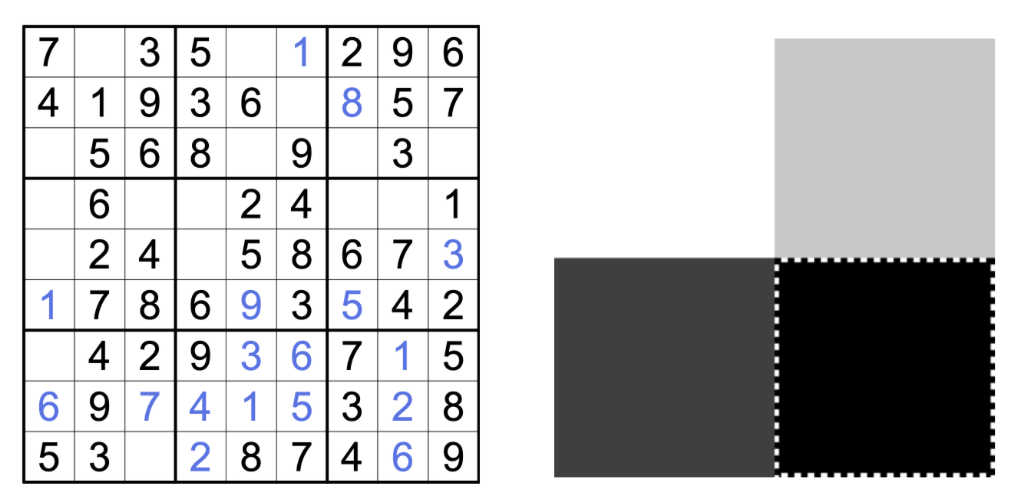}
    \caption{Example of a semi-completed Sudoku board and the corresponding heatmap visualization.}
    \label{fig:exSudokuVis}
    \vspace{-1em}
\end{figure}

For Sudoku, the maximum corresponded to the most completed quadrant, calculated as the percentage of filled cells relative to its initial empty cells, with shared central cells contributing to adjacent quadrants. This value was visualized as the darkest region in the heatmap (Fig.~\ref{fig:exSudokuVis}).

The secondary perception task was dynamically triggered when a relevant point in the games was reached: 25\% of the board completed in Sudoku and a new maximum possible piece-length sequence in Connect 4. By triggering the secondary task in these moments, we were able to evaluate whether contextual timing resulted in higher comprehension accuracy or lower disruption compared to other patterns.

To reduce sensory interference, prompts were delivered via text-to-speech audio, and responses were input using a physical numpad mapped spatially to the visualization layout (four keys for bar charts; 2x2 for heatmaps).
Response time and accuracy were logged for each trial. Additionally, an ambiguity factor quantified the relative difference between the two most prominent values:

\[
\text{AF}_{\text{Connect Four}} = \frac{|v_1 - v_2|}{7}, \quad
\text{AF}_{\text{Sudoku}} = \frac{|v_1 - v_2|}{100}.
\]

The ambiguity factor captured the limits of pre-attentive perception by quantifying how distinct the two highest values were. Pre-attentive processing is most effective when value differences are clear. However, when values are numerically close, distinguishing the maximum becomes perceptually difficult. In such cases, the difference may fall below the just-noticeable difference threshold, reducing accuracy even with correct task execution. 
This factor was applied in cases of near-equal alternatives (similarity$>75\%$) to distinguish genuine performance errors from perceptual limits.

The Meta Quest 3 headset was set in pass-through AR mode for stereoscopic rendering and integrated tracking, with built-in audio delivering task prompts. Selected sessions were recorded using the headset’s built-in video capture for validation and analysis.

\begin{figure}[t]
  \centering
  \begin{subfigure}[b]{0.4\textwidth}
    \includegraphics[width=\textwidth]{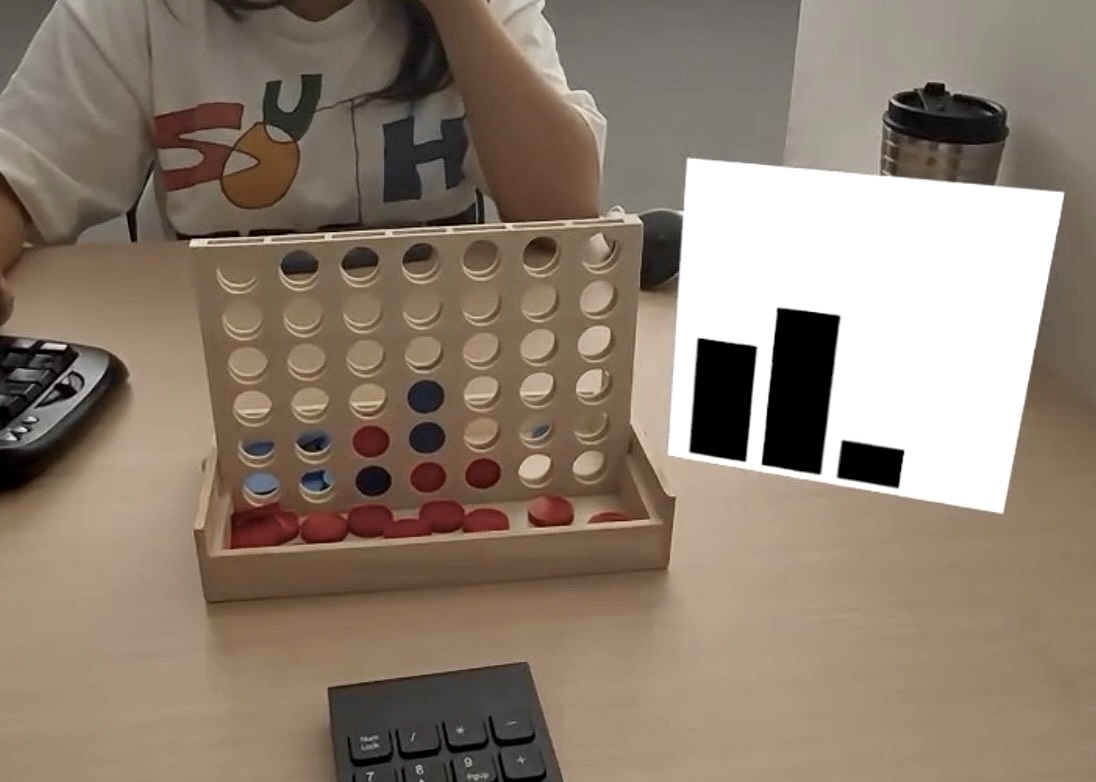}
    \caption{Placement of the barchart visualization.}
    \label{fig:exampleC4}
  \end{subfigure}
  \hfill
  \begin{subfigure}[b]{.4\textwidth}
    \includegraphics[width=\textwidth]{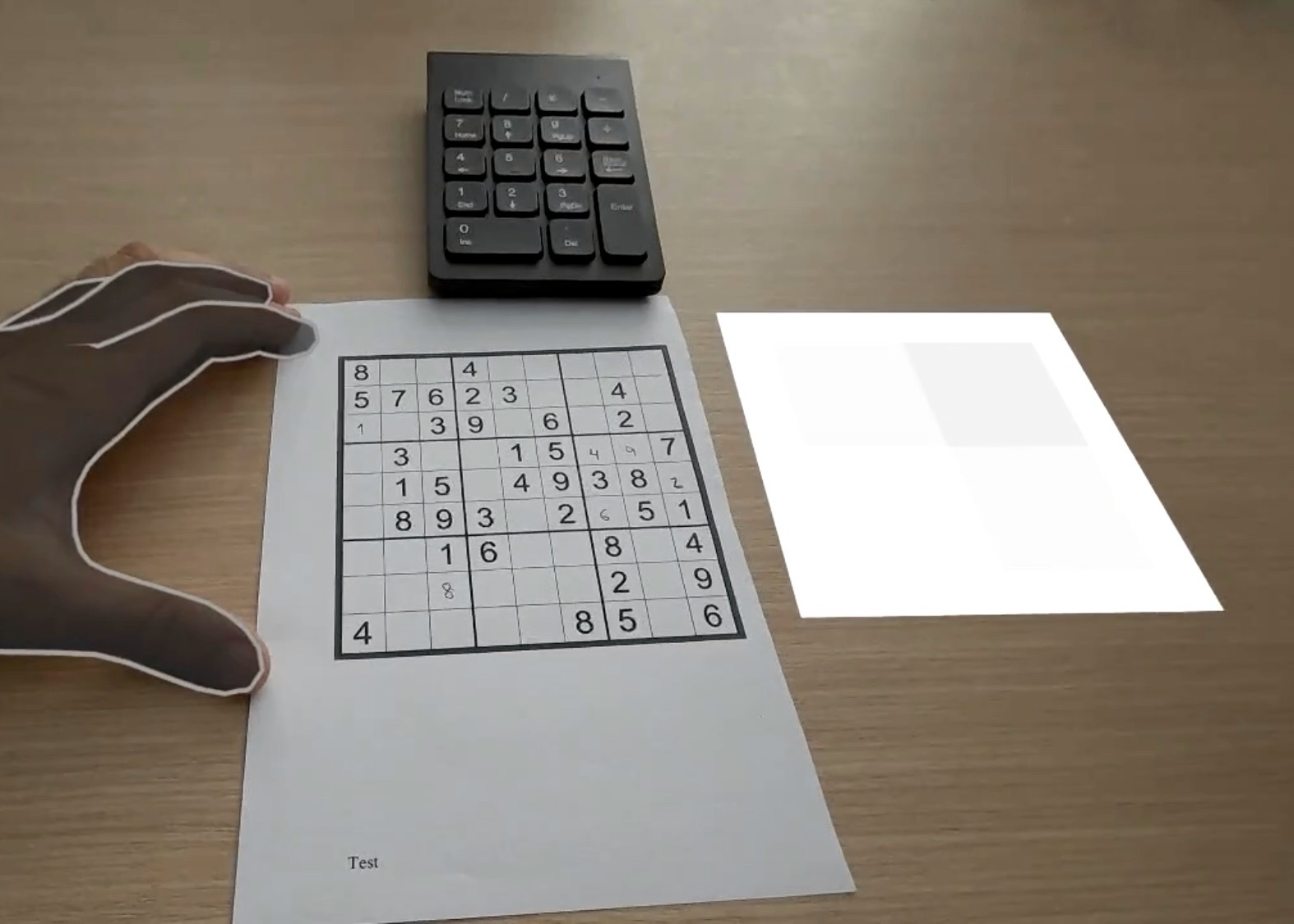}
    \caption{Placement of the heatmap visualization.}
    \label{fig:exampleSK}
  \end{subfigure}
  \caption{Placement of the visualizations used in the study: the barchart (left) and the heatmap (right).}
  \label{fig:placementExamples}
  \vspace{-1em}
\end{figure}

Following established design guidelines for AR interfaces \cite{Lee:2024, Imamov:2020, Piening:2021}, the visualizations were presented in a world-fixed AR panel positioned near the game board, in the bottom-right corner of the user’s field of view. This placement avoided occlusion of the primary task and ensured proximity to the task space while minimizing occlusion and maintaining peripheral visibility. The panel was aligned perpendicular to the user’s point of view, ensuring comfortable readability while supporting quick glance-based access, as recommended in prior work on interface positioning and task-switching efficiency \cite{Imamov:2020}.  
An example of the visualization placement used in the study is shown in Figure~\ref{fig:placementExamples}.

\subsubsection{Study Variables}
Our study had three dependent variables. \textbf{Comprehension accuracy} measured participants’ ability to correctly retrieve information from visualizations, independent of primary task performance. For Connect Four, this entailed selecting the highest bar in the bar chart; for Sudoku, identifying the heatmap region with the highest intensity. Accuracy was complemented by response time and an ambiguity factor, which accounted for reduced pre-attentive processing when the two highest values were close. \textbf{Disruption level} captured how much the visualization interrupted the primary task, assessed via a 5-point Likert scale and supplemented by mental workload using an unweighted NASA-TLX questionnaire \cite{Hart:1988}. Finally, \textbf{task performance} evaluated the impact of visualizations on the main task, using game-specific metrics: for Connect Four, outcomes (win/loss) combined with duration (quicker victories and slower losses indicate better performance); for Sudoku, normalized final completion percentage.

The single within-subject independent variable was \textbf{visualization type}, with four visualization patterns differing in frequency and timing.

\subsubsection{Participants}
Participants were limited to young adults (20 male, 10 female) aged 18–24 years with engineering backgrounds to ensure a homogeneous group. 

Gender diversity was considered within volunteer possibilities to ensure broader representation. Sixty percent reported prior exposure to AR, while 27\% had none and 13\% were unsure. Self-rated familiarity on a 1–5 scale decreased from VR ($M = 3.0$, $SD = 0.96$) to AR headsets ($M = 2.0$, $SD = 1.07$) and interaction design/human computer interaction ($M = 1.0$, $SD = 1.08$), indicating moderate technological familiarity but limited design expertise. All participants had normal or corrected-to-normal vision, with the setup accommodating glasses and contact lenses. Participation was voluntary, and informed consent was obtained in accordance with institutional ethics guidelines.

\subsubsection{Study Procedure}
The study was conducted in person with 30 participants in a controlled laboratory environment, structured into three main phases: an initial screening phase, a preliminary testing phase, and a subsequent experimental phase.

In the screening phase, participants completed a background questionnaire gathering demographic information as well as prior exposure to AR-related technologies. 
The first page of the questionnaire also explained data logging and optional video recording of headset views for analysis. Written consent was obtained from all participants, preventing non-consensual data collection. 

Because IVs are intended to familiarly blend into daily life \cite{Moreira:2024, Moreira:2023a}, yet participants are unlikely to have prior experience with these visualizations, a preliminary phase was required to ensure participants could correctly understand the visualizations and the information therein.

To that end, participants received a brief overview in the preliminary phase outlining the study’s objectives and structure, introducing the two visualizations (bar chart and heatmap) through printed examples. To ensure correct interpretation, they were asked to identify the higher value in each visualization type. Participants then set up the Meta Quest 3 headset and a disposable sweat guard to ensure hygiene and comfort.
Then, they performed two warm-up sessions as trial runs of each task before the main experiments, allowing users to familiarize themselves with gameplay and the AR interaction. During these warm-ups, the visualization was shown in ambient mode to familiarize participants with the information layout.
Participants could repeat warm-ups until they felt confident with the interaction and task flow. 

The testing phase began with each participant completing four iterations of each of the two primary tasks (Sudoku and Connect Four) for each AR visualization pattern (Ambient, Incidental, Periodic, and None). The order of tasks, visualization conditions, and equivalent Sudoku boards were counterbalanced across participants using a Latin Square design to reduce order and learning effects. The Sudokus were presented using A5 paper puzzles, while the Connect Four was played on a physical wooden board with red and blue game pieces (Figure~\ref{fig:placementExamples}).

\subsection{Software Implementation}
Custom Python programs managed both game tasks, automated visualization rendering, event logging, and communication with Unity running on the Meta Quest headset. Logging captured all game states, participant moves, visualization triggers, and responses with timestamps in structured CSV files. Digital questionnaires (background, post-trial, post-study) were delivered to participants via Google Forms. 

\paragraph{Connect Four}
The 6×7 board was implemented in Python using \texttt{pygame}, and was represented with \texttt{NumPy} arrays. The computer opponent used a Minimax algorithm with Alpha-Beta pruning (depth 3) \cite{JustA3DObject:2023}, with a 40\% random-move probability after the first five turns to simulate human-like variability. This computer agent acted as a "standardized opponent," ensuring that every participant faced the same level of strategic difficulty while still experiencing the unpredictability of a real-world game. The system handled move validation, win detection, turn alternation, visualization triggers, and automatic logging of all actions, responses, and outcomes.

\paragraph{Sudoku}
The sudoku game was implemented in Python with a Flask (v3.1.0) backend, and its grids were divided into four quadrants, tracking initial, updated, and completed states. The four puzzles (plus trial) were of medium difficulty (38 clues, unique solution) and generated through equivalence transformations to maintain difficulty across iterations. The system managed visualization triggers, secondary task timing, and logged all participant interactions automatically. Each iteration was limited to 4 minutes, allowing sufficient opportunities to trigger the secondary accuracy task at each 25\% quadrant completion and prevent participant fatigue.

%% file: sections/results.tex
\section{Results}
The following section presents the experimental results across all dependent variables. 
Data was screened for outliers, and all extreme outliers (points beyond $3 \times IQR$) were removed. However, we kept them in the boxplots (shown below) to give the original overview. Data was generally not normally distributed, as assessed by Shapiro–Wilk’s test ($p < .05$). Therefore, non-parametric Friedman tests were used on all dependent variables with variance to determine statistically significant differences
with Bonferroni-corrected post hoc pairwise comparisons when appropriate \cite{LaerdStatistics2015}, as an alternative to the one-way repeated measures ANOVA. Groups with zero variance were excluded from testing, as they could not be meaningfully compared in non-parametric tests. All analyses were conducted using IBM SPSS Statistics (SPSS Statistics, 2012).
Each subsection presents the statistics, boxplot visualizations, and a final summary of significant differences \ref{tab:summarySignificantDifferences}.

\subsection{Accuracy}
The following analyses examine accuracy in the secondary task across the four visualization conditions for both tasks. 
\subsubsection{Accuracy with Ambiguity Measure}
This subsection presents the accuracy results incorporating the ambiguity factor for cases where the visualization values were closely similar and distinguishing between clearly perceivable and ambiguous visualization conditions.

\begin{figure}[t]
  \centering
  \begin{subfigure}[b]{0.23\textwidth}
    \includegraphics[width=\textwidth]{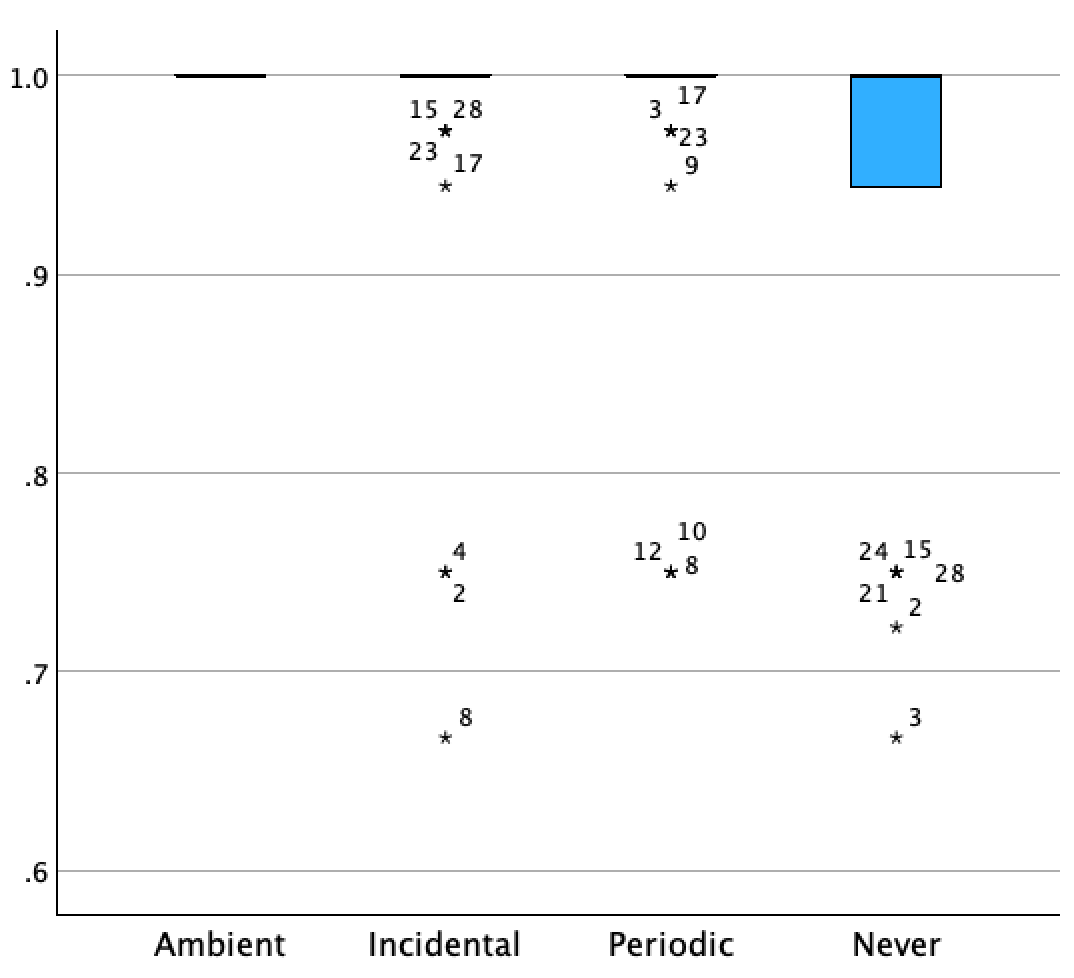}
    \caption{Connect Four Accuracy with Ambiguity Measure}
    \label{fig:c4AccuracyBoxplot}
  \end{subfigure}
  \hfill
  \begin{subfigure}[b]{.23\textwidth}
    \includegraphics[width=\textwidth]{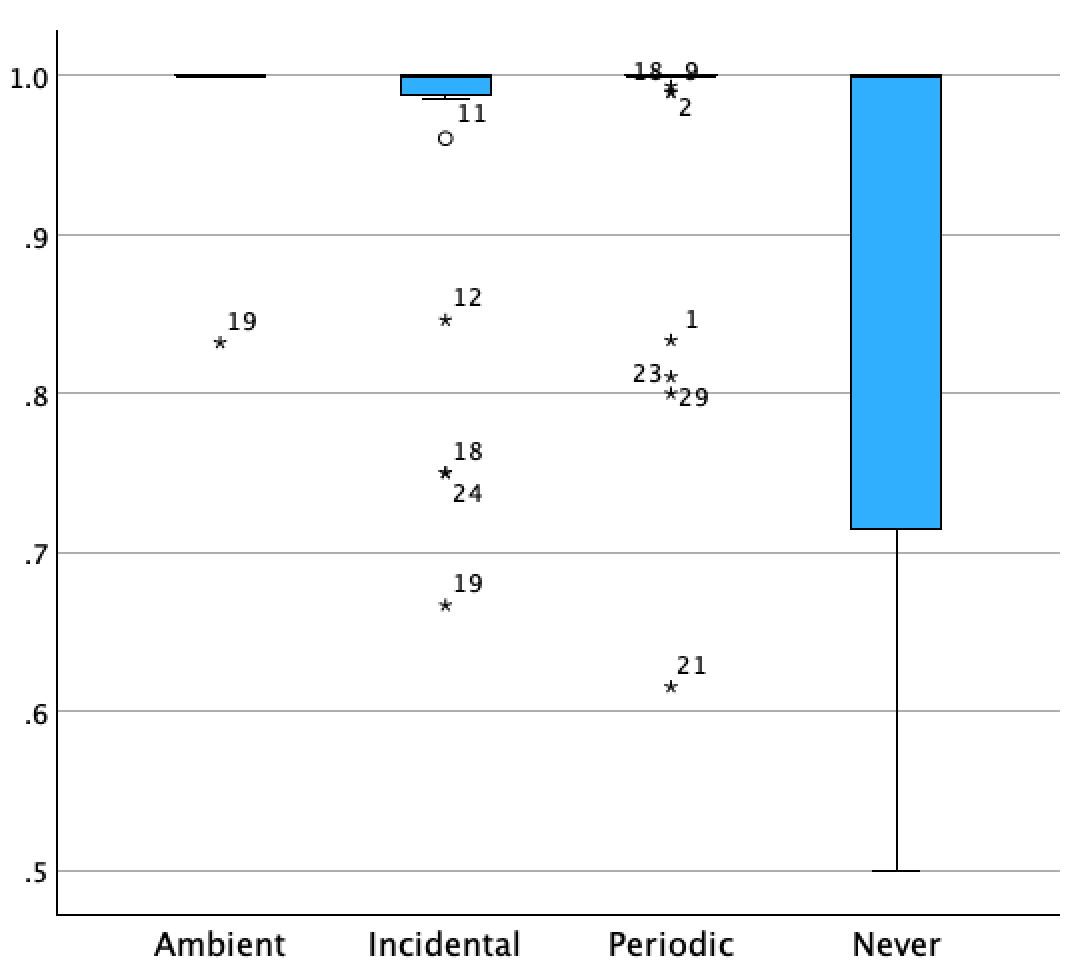}
    \caption{Sudoku Accuracy with Ambiguity Measure}
    \label{fig:skAccuracyBoxplot}
  \end{subfigure}
  \caption{Accuracy scores with the ambiguity measure across visualization conditions for Connect Four and Sudoku tasks.}
  \label{fig:boxplotAccuracy}
  \vspace{-1em}
\end{figure}

\paragraph{Connect Four}
All ambient, incidental, and periodic modes presented no within-group variance and were excluded from the statistical test. Because comparing a group with variance to one without will always yield statistically significant results, we concluded there were significant differences between None and all other modes.
The distribution of accuracy scores for this task, across visualization conditions, is shown in Figure~\ref{fig:c4AccuracyBoxplot}.

\paragraph{Sudoku}
In Sudoku, the ambient and periodic modes had no variance after outlier removal. The Friedman test was conducted on the two remaining groups, with a significant difference between incidental ($Mdn = 1.00, Avg = .964$) and None ($Mdn = 1.00, Avg = 0.869$) ($p = .00
25$). All pairwise comparisons between the groups with and without variance were also deemed significant, between ambient and periodic and between both incidental and periodic. 
The distribution of accuracy scores for this task and across visualization conditions is shown in Figure~\ref{fig:boxplotAccuracy}.

\subsubsection{Accuracy with Time Measure}
This subsection presents the accuracy results weighted by response time, reflecting efficiency in interpreting visual information, where faster, correct responses indicate stronger perceptual understanding, and slower, incorrect responses indicate higher difficulty or uncertainty.
For this dataset, the minimum and maximum values used for score normalization were considered the second lowest and second highest values. The first were considered outliers (.000 seconds and over 78 seconds due to misclicks) and not included in the scores.

\begin{figure}[t]
  \centering
  \begin{subfigure}[b]{0.23\textwidth}
    \includegraphics[width=\textwidth]{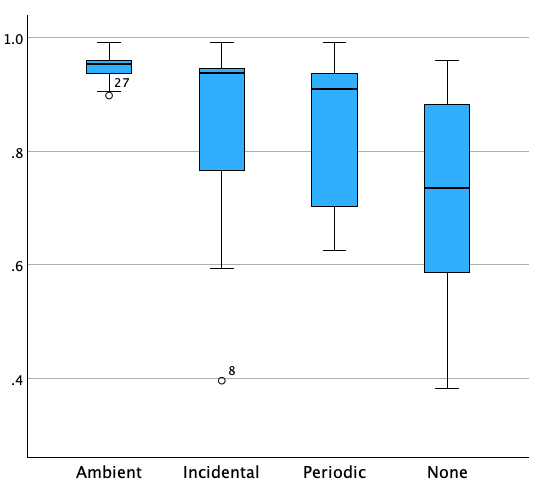}
    \caption{Connect Four Accuracy with Time Measure}
    \label{fig:c4AccTimeBoxplot}
  \end{subfigure}
  \hfill
  \begin{subfigure}[b]{0.23\textwidth}
    \includegraphics[width=\textwidth]{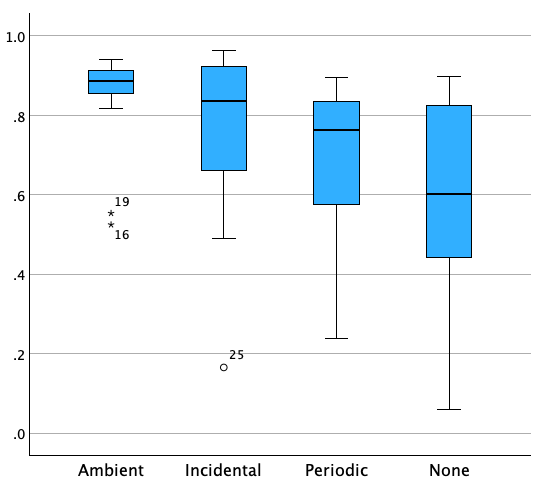}
    \caption{Sudoku Accuracy with Time Measure}
    \label{fig:skAccTimeBoxplot}
  \end{subfigure}
  \caption{Accuracy scores with the time measure across visualization conditions for Connect Four and Sudoku tasks.}
  \label{fig:boxplotAccuracyTime}
  \vspace{-1em}
\end{figure}

\paragraph{Connect Four}
The accuracy scores were statistically significantly different for the visualization modes $\chi^2(2) = 38.880, p < .001$. Post hoc analysis revealed statistically significant differences between None ($Mdn = 0.736$) and all other patterns, with ambient ($Mdn = 0.953$) ($p < .001$), incidental ($Mdn = 0.937$) ($p = .002$) and periodic ($Mdn = 0.910$) ($p = .016$). 
The only other significant increase was found between ambient ($Mdn = 0.953$) and periodic ($Mdn = 0.910$) ($p = .008$).
The distribution of accuracy scores for this task and across visualization conditions is shown in Figure~\ref{fig:skAccTimeBoxplot}.
\paragraph{Sudoku}
The accuracy scores were statistically significantly different for the visualization modes $\chi^2(2) = 30.840, p < .001$. Post hoc analysis revealed statistically significant differences between None ($Mdn = .601$) and incidental ($Mdn = .836$) ($p = .001$), also between None ($Mdn = .601$) and ambient ($Mdn = .886$) ($p < .001$). Furthermore, there was also a significant decrease from ambient ($Mdn = .886$) to periodic ($Mdn = .763$) ($p < .001$). 
The distribution of accuracy scores for this task and across visualization conditions is shown in Figure~\ref{fig:boxplotAccuracyTime}.

\subsection{Primary Task Disruption}
This section examines the impact of different visualization modes on participants’ experience during the primary task, considering both perceived disruption and mental workload. The None condition displayed no variance for both tasks and was excluded from statistical testing, considering all pairwise comparisons between None and the other modes significant. 
\subsubsection{Perceived Disruption}
\begin{figure}[t]
  \centering
  \begin{subfigure}[b]{0.23\textwidth}
    \includegraphics[width=\textwidth]{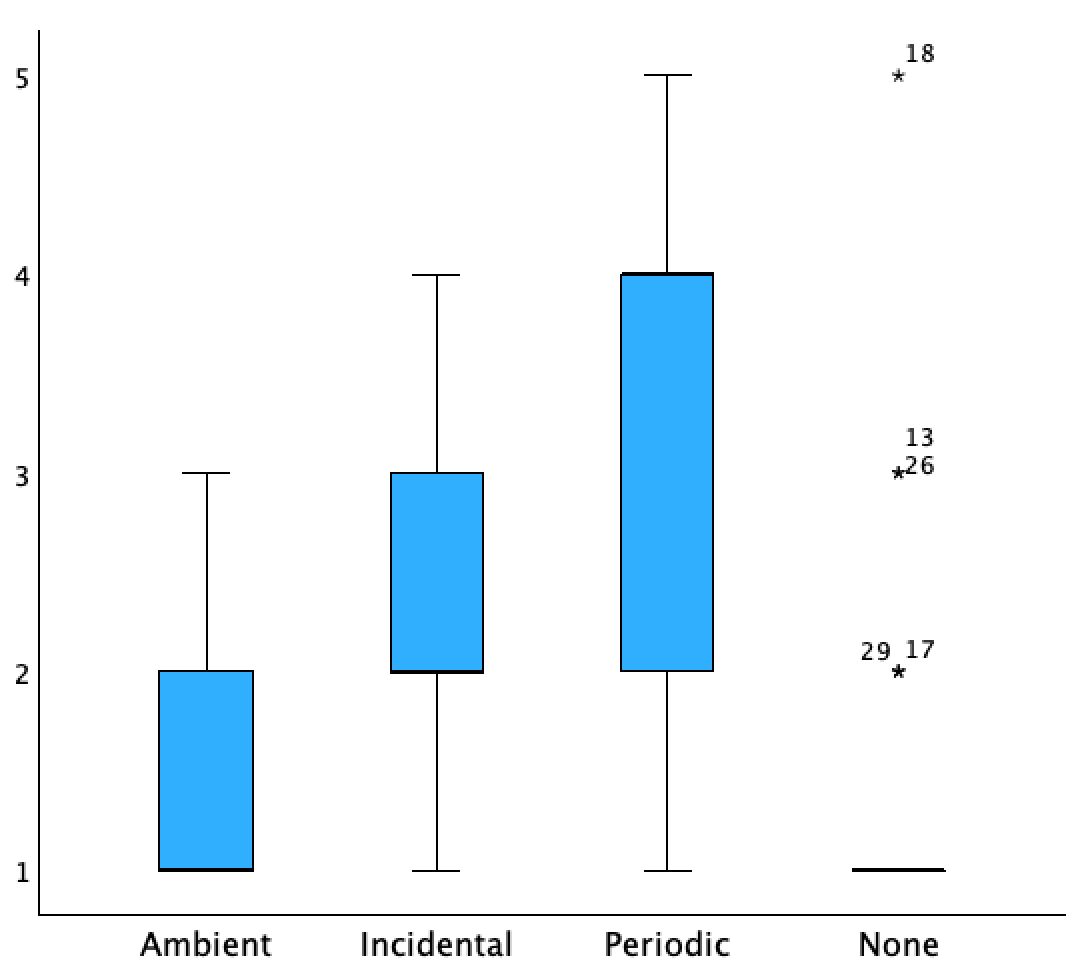}
    \caption{Connect Four Disruption}
    \label{fig:c4DisruptBoxplot}
  \end{subfigure}
  \hfill
  \begin{subfigure}[b]{.23\textwidth}
    \includegraphics[width=\textwidth]{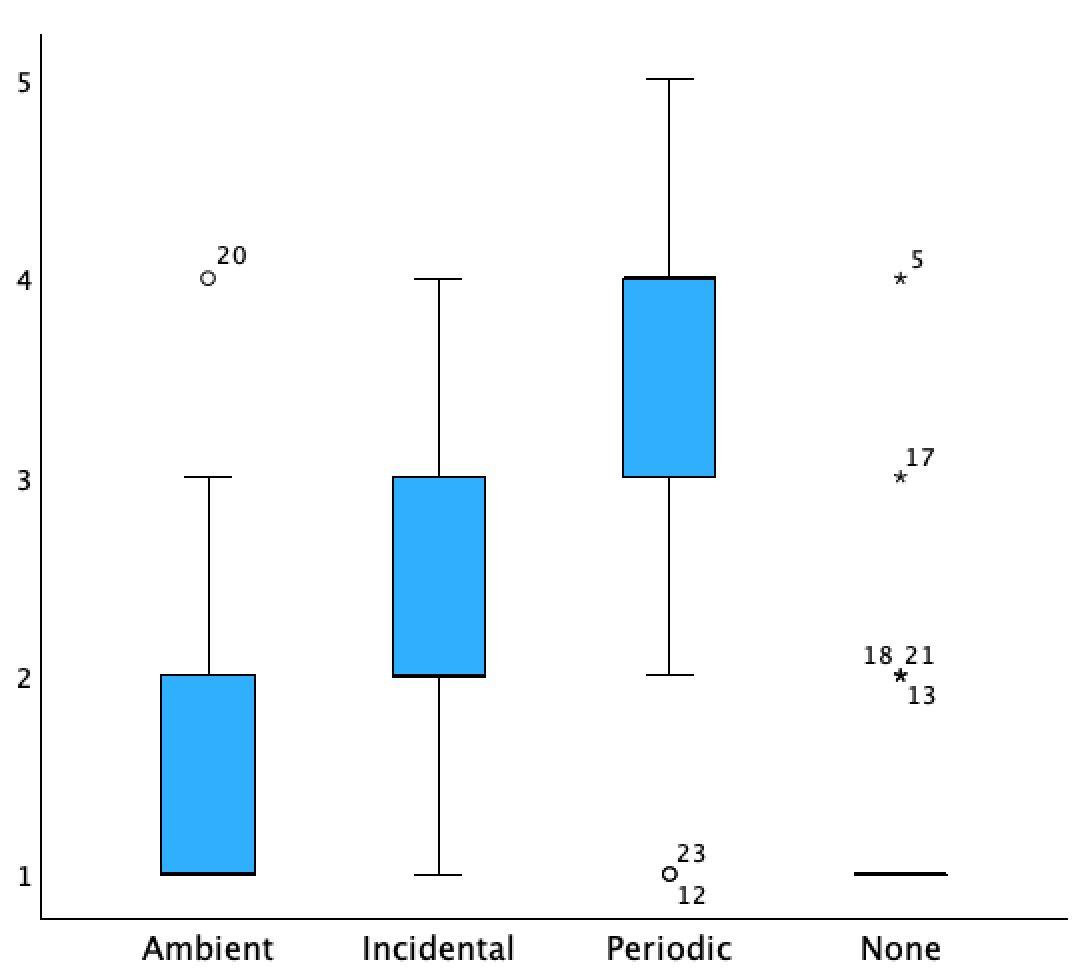}
    \caption{Sudoku Disruption}
    \label{fig:skDisruptBoxplot}
  \end{subfigure}
  \caption{Disruption levels (Likert scale) across visualization modes for both tasks.}
  \label{fig:boxplotDisruption}
  \vspace{-1em}
\end{figure}

First, we present the results for perceived disruption.
\paragraph{Connect Four}
Between the remaining conditions (ambient, incidental, and periodic), the Friedman test revealed significant differences ($p < .001$). Post hoc analysis showed lower perceived disruption for ambient ($Mdn = 0.953$) compared to both periodic ($Mdn = 0.910$) ($p < .001$) and incidental ($Mdn = 0.937$) ($p = .003$). 
Figure~\ref{fig:c4DisruptBoxplot} presents the results of the disruption measure for this task across the visualization conditions.

\paragraph{Sudoku}
The accuracy scores were statistically significantly different for the visualization modes $\chi^2(2) = 34.865, p < .001$. Post hoc analysis revealed statistically significant differences between ambient ($Mdn = .886$) and incidental ($Mdn = .836$) ($p = .001$) and also between ambient and periodic ($Mdn = .763$) ($p < .001$). Furthermore, there was also a significant increase from incidental to periodic ($p < .001$). 
Figure~\ref{fig:skDisruptBoxplot} presents the results of disruption for this task across the visualization conditions.

\subsubsection{Perceived Workload}
\begin{figure}[t]
  \centering
  \begin{subfigure}[b]{0.23\textwidth}
    \includegraphics[width=\textwidth]{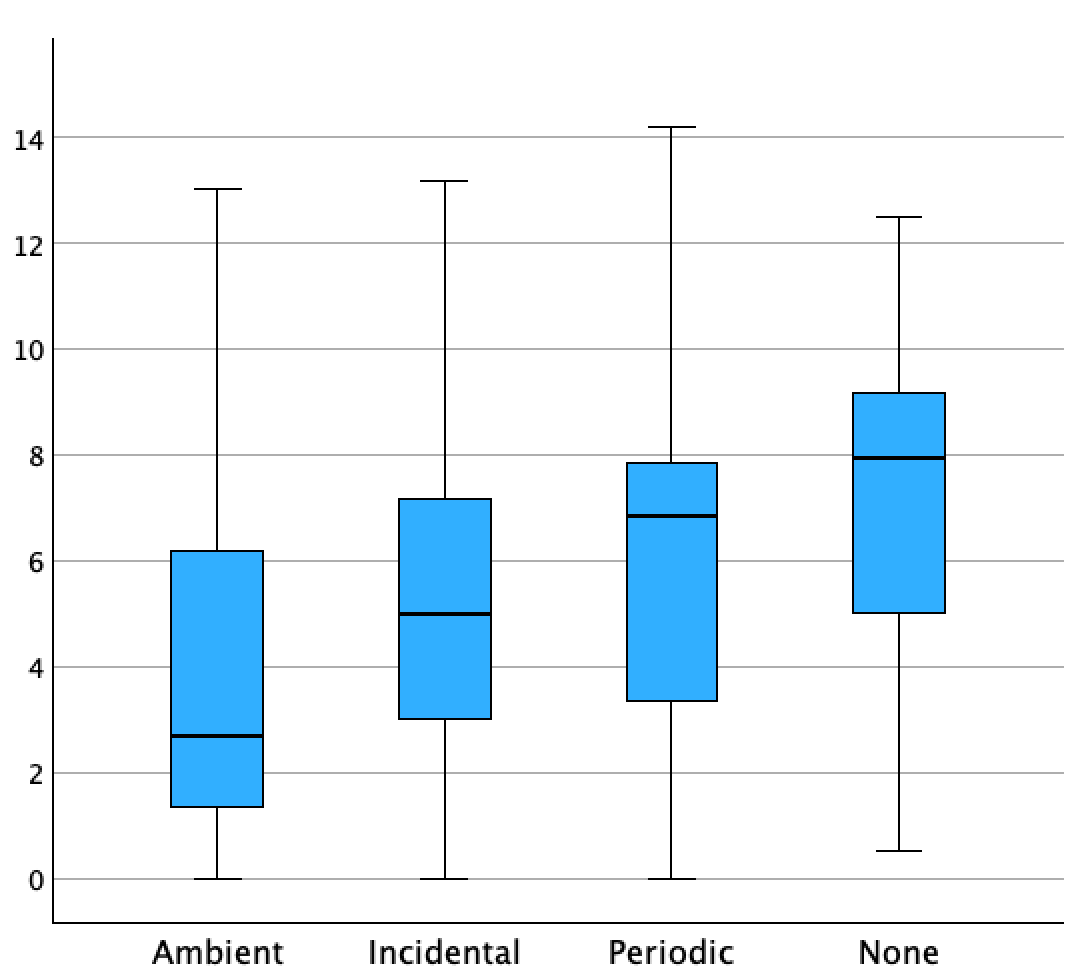}
    \caption{Connect Four Workload}
    \label{fig:c4NASABoxplot}
  \end{subfigure}
  \hfill
  \begin{subfigure}[b]{.23\textwidth}
    \includegraphics[width=\textwidth]{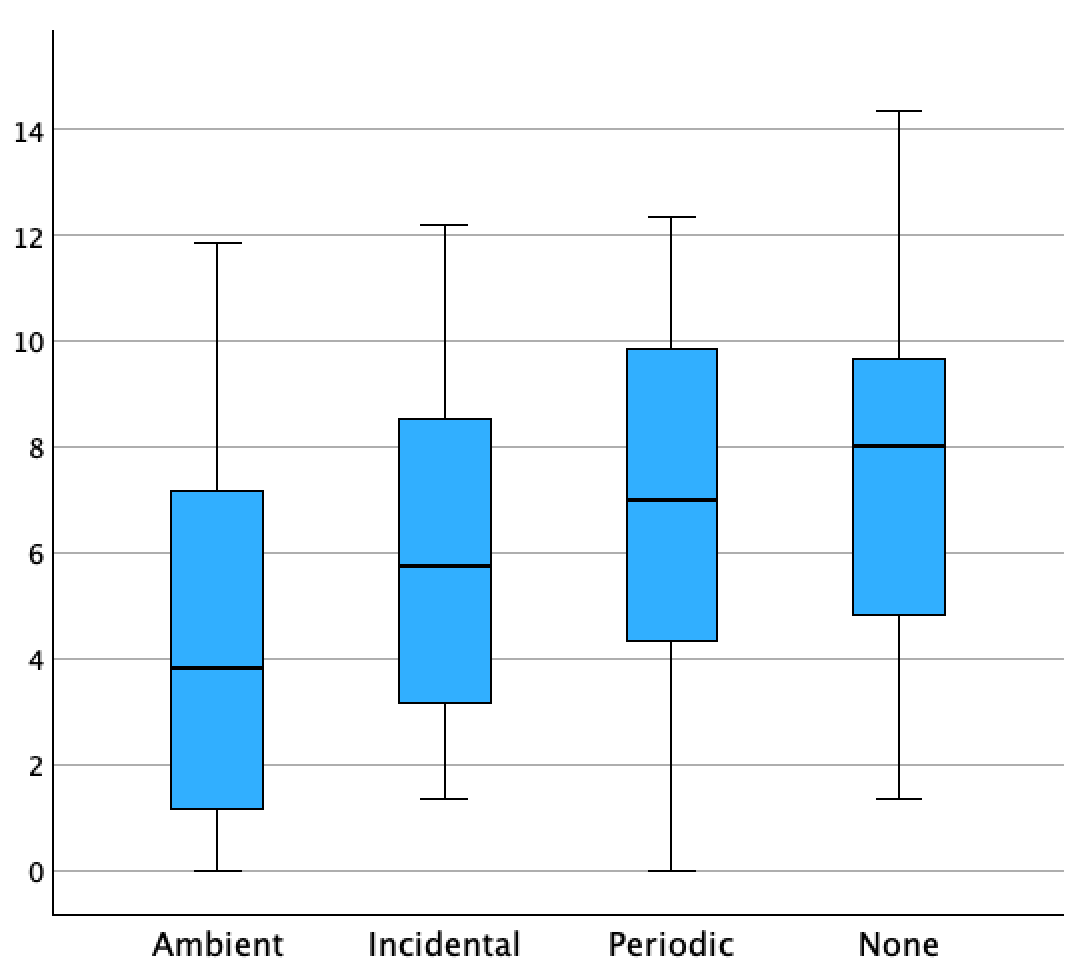}
    \caption{Sudoku Workload}
    \label{fig:skNASABoxplot}
  \end{subfigure}
  \caption{NASA-TLX mental workload scores across visualization modes for both tasks.}
  \label{fig:boxplotNASATLX}
  \vspace{-1em}
\end{figure}

Now, we present the results for perceived workload using the NASA-TLX.
\paragraph{Connect Four}
The resulting NASA-TLX workload scores were statistically significantly different at the different time points during participantion ($\chi^2(2) = 28.252, p < .001$). Post hoc analysis revealed statistically significant differences in workload scores from ambient ($Mdn = 2.667$) to periodic ($Mdn = 6.833$) ($p = .031$), ambient to None ($Mdn = 7.917$) ($p < .001$) and incidental ($Mdn = 5.000$) ($p = .003$) to None, but not incidental to ambient, incidental to periodic, or periodic to None.
The distribution of NASA-TLX workload evaluation results across visualization conditions is shown in Figure~\ref{fig:c4NASABoxplot}.

\paragraph{Sudoku}
The resulting NASA-TLX workload scores were statistically significantly different at the different time points during the exercise intervention ($\chi^2(2) = 29.868, p < .001$). Post hoc analysis revealed statistically significant differences in workload scores from ambient ($Mdn = 3.833$) to periodic ($Mdn = 7.000$) ($p < .001$), ambient to None ($Mdn = 8.000$) ($p < .001$) and incidental ($Mdn = 5.750$) ($p = .008$) to None, but not incidental to ambient, incidental to periodic, or periodic to None.
The distribution of NASA-TLX workload evaluation results across visualization conditions is shown in Figure~\ref{fig:skNASABoxplot}.

\subsection{Primary Task Performance}
Now we present the results regarding the performance of each of the primary tasks tested.

\begin{figure}[b]
  \centering
  \begin{subfigure}[b]{0.23\textwidth}
    \includegraphics[width=\textwidth]{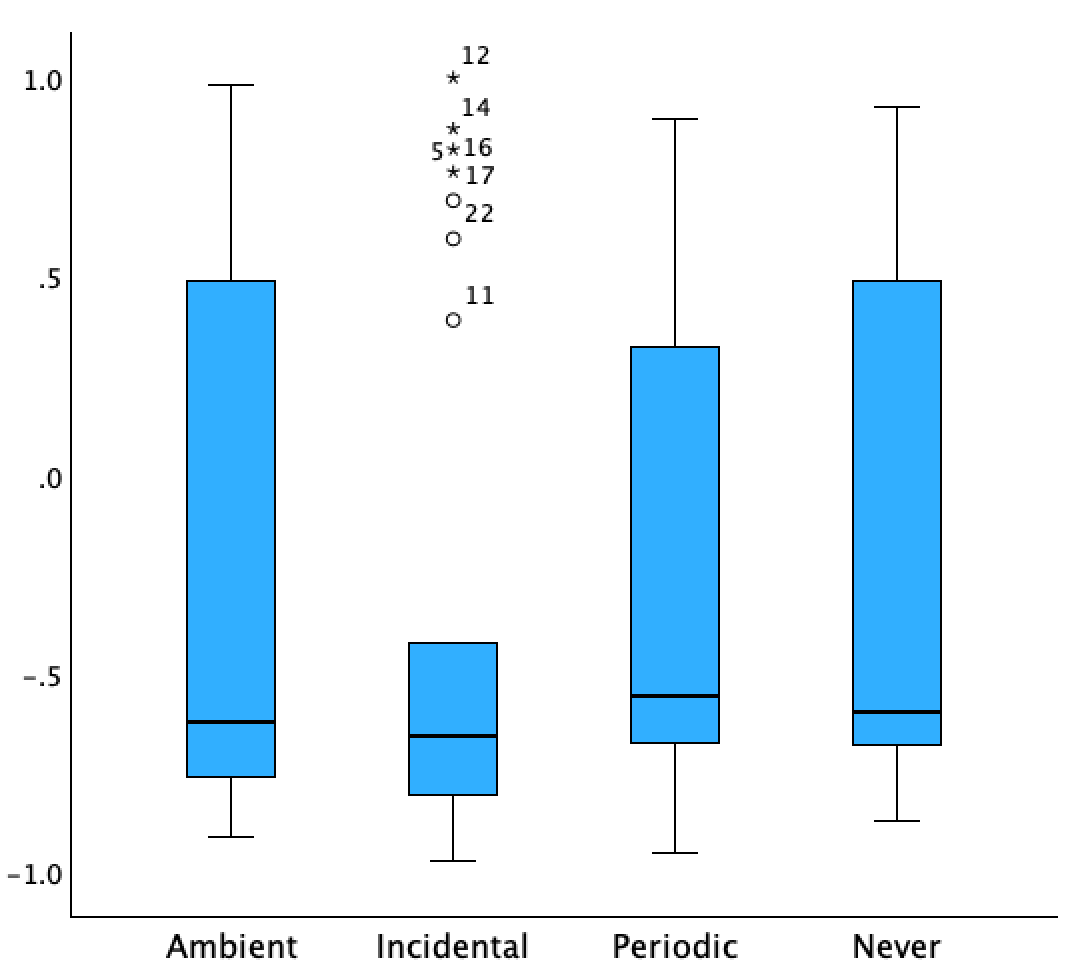}
    \caption{Connect Four Performance}
    \label{fig:c4PerformanceBoxplot}
  \end{subfigure}
  \hfill
  \begin{subfigure}[b]{.23\textwidth}
    \includegraphics[width=\textwidth]{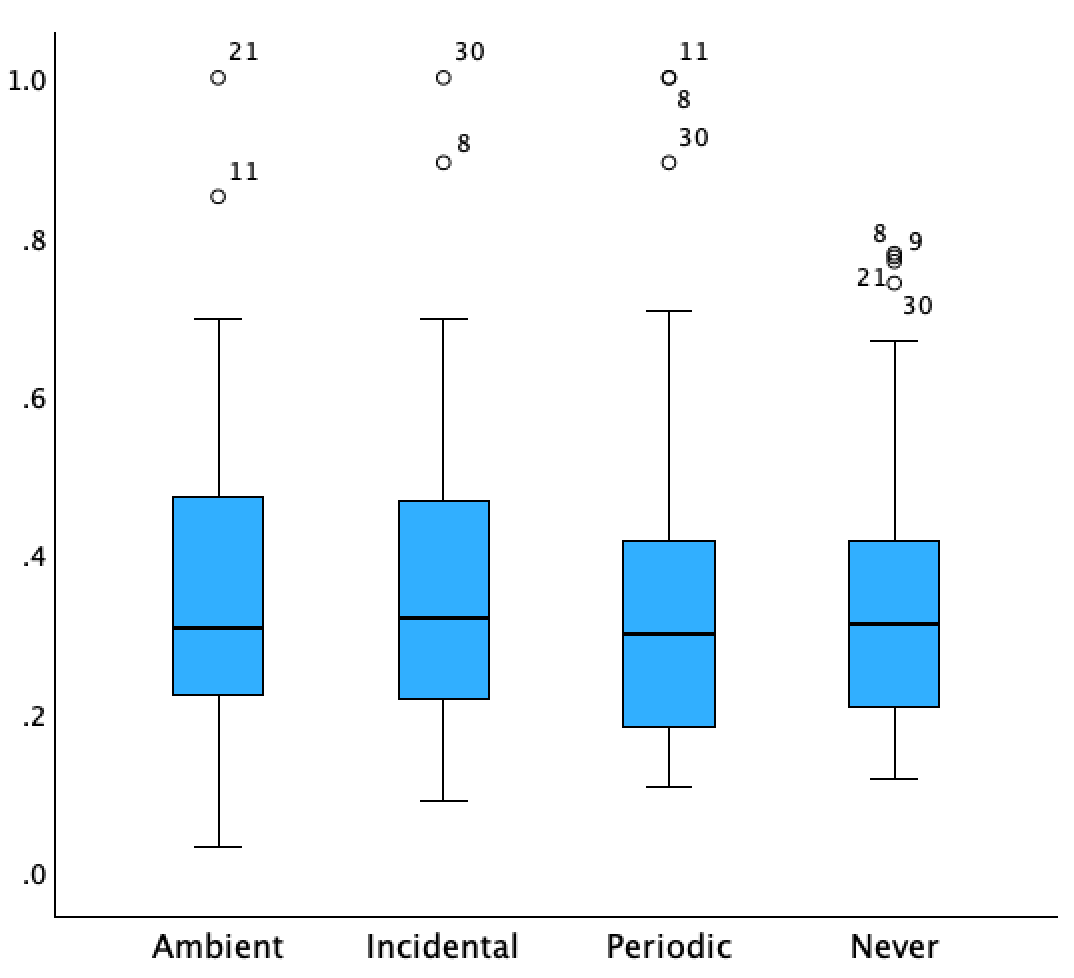}
    \caption{Sudoku Performance}
    \label{fig:skPerformanceBoxplot}
  \end{subfigure}
  \caption{Performance results across visualization modes for both tasks.}
  \label{fig:boxplotPerformance}
  \vspace{-1em}
\end{figure}

\paragraph{Connect Four}
The obtained scores were ($Mdn = -0.657$) for incidental, ambient ($Mdn = -0.620$), None ($Mdn = -0.598$) and periodic ($Mdn = -0.554$), but results from the statistical test revealed no significant differences ($\chi^2(2) = 5.217, p = .157$).
The performance results for Connect Four are presented in Figure~\ref{fig:c4PerformanceBoxplot}.

\paragraph{Sudoku}
The scores varied from incidental ($Mdn = 0.321$), to None ($Mdn = 0.313$), to ambient ($Mdn = 0.306$) and periodic ($Mdn = 0.300$), but the differences were not statistically significant ($\chi^2(2) = 2.636, p = .451$).
The performance results for Sudoku are presented in Figure~\ref{fig:skPerformanceBoxplot}.

\noindent Table~\ref{tab:summarySignificantDifferences} summarizes all statistically significant pairwise differences found across the analyzed dependent variables. Only comparisons with $p < .05$ after Bonferroni correction are reported. 
No significant effects were observed for performance, and accuracy differences were limited to the Sudoku task.

\begin{table}[ht]
\centering
\small % reduce font size to help fit text
\caption{Summary of significant pairwise differences across dependent variables}
\label{tab:summarySignificantDifferences}
\arrayrulecolor{lightgray} % set line color
\renewcommand{\arraystretch}{1.2} % row height
\begin{tabular}{|p{1.8cm}|p{.8cm}|p{2.8cm}|p{1.5cm}|}
\hline
\textbf{\makecell[l]{Dependent \\ Variable}} & \textbf{Task} & \textbf{Comparison} & \textbf{Significance} \\
\hline
% ----------------- ACCURACY WITH AMBIGUITY -----------------
\multirow{4}{*}{\makecell[l]{Accuracy with \\ Ambiguity}} 
    & C4 & None vs Amb, Inc, Per & implicit$^\dagger$ \\
    \cline{2-4}
    & SK & ambient vs Inc, None & implicit$^\dagger$ \\
    & SK & periodic vs Inc, None & implicit$^\dagger$ \\
    & SK & incidental vs None & $p = .002$ \\[3pt] \hline

% ----------------- ACCURACY WITH TIME -----------------
\multirow{7}{*}{\makecell[l]{Accuracy \\ with Time}} 
    & C4 & None vs ambient & $p < .001$ \\
    & C4 & None vs incidental & $p = .002$ \\
    & C4 & None vs periodic & $p = .016$ \\
    & C4 & ambient vs periodic & $p = .008$ \\
    \cline{2-4}
    & SK & None vs incidental & $p = .001$ \\
    & SK & None vs ambient & $p < .001$ \\
    & SK & ambient vs periodic & $p < .001$ \\[3pt] \hline

% ----------------- DISRUPTION -----------------
\multirow{7}{*}{\makecell[l]{Disruption \\ (Likert Scale)}} 
    & C4 & None vs Amb, Inc, Per & implicit$^\dagger$ \\
    & C4 & ambient vs incidental & $p = .003$ \\
    & C4 & ambient vs periodic & $p < .001$ \\
    \cline{2-4}
    & SK & None vs Amb, Inc, Per & implicit$^\dagger$ \\
    & SK & ambient vs incidental & $p = .001$ \\
    & SK & ambient vs periodic & $p < .001$ \\
    & SK & incidental vs periodic & $p < .001$ \\[3pt] \hline

% ----------------- WORKLOAD -----------------
\multirow{6}{*}{\makecell[l]{Workload \\ (NASA-TLX)}} 
    & C4 & ambient vs periodic & $p = .031$ \\
    & C4 & ambient vs None & $p < .001$ \\
    & C4 & incidental vs None & $p = .003$ \\
    \cline{2-4}
    & SK & ambient vs periodic & $p < .001$ \\
    & SK & ambient vs None & $p < .001$ \\
    & SK & incidental vs None & $p = .008$ \\[3pt] \hline

% ----------------- PERFORMANCE -----------------
\multirow{1}{*}{Performance} 
    & Both & – & n.s.\\
\hline
\end{tabular}
\vspace{0.3em}
\parbox{.95\linewidth}{
    {\footnotesize \textit{Note.} C4 = Connect Four; SK = Sudoku; Amb = ambient; Inc = incidental; Per = periodic.\\
    $^\dagger$Zero-variance groups excluded; any comparison with a group with variance is inherently significant.}
}
\vspace{-1em}
\end{table}

\subsection{Discussion}
Accuracy results demonstrated a significant improvement from None to all other patterns. However, no additional differences were observed, and all except one were implicit conclusions, likely attributable to ceiling effects from the chosen tasks.
Time-based accuracy revealed significant differences between None and all visualization modes in Connect Four, as well as between None and both ambient and incidental modes in Sudoku, indicating faster information acquisition with visual support. Thus, \textbf{H1 is supported}. 
The incidental mode did not outperform periodic (\textbf{H2 not supported}) but achieved comparable performance to ambient, while periodic did not, in time-based accuracy, suggesting that contextual, transient cues can aid comprehension. Therefore, \textbf{H3 is supported}, confirming that IVs can be as effective as ambient visualizations.

Although None provided the highest workload scores, requiring higher mental demand for the task due to the lack of additional information, as predicted, the absence of visualization (None) caused the least disruption (\textbf{H4 is supported}). Ambient visualizations were perceived as the next least intrusive, while periodic and incidental modes showed higher disruption. 
NASA-TLX results confirmed this trend, with ambient providing the lowest workload and disruption, followed by incidental and then periodic. 
Although \textbf{H5 was not supported}, incidental and ambient showed similar disruption levels, while the time-based mode differed significantly, further suggesting that context-triggered visualizations are less distracting than time-based ones. 

Visualization modes did not significantly affect primary task outcomes, indicating that added visualizations did not hinder gameplay. This \textbf{supports H6}, showing that IVs and other visualization types do not disrupt task performance. However, \textbf{H7 was not supported}, as the incidental pattern did not outperform other modes but maintained equivalent performance, underscoring its suitability for AR.

Although comparing tasks was not a primary aim, their differing cognitive demands produced distinct accuracy and disruption patterns. For ambiguity-based accuracy, Sudoku showed contrasts, with ambient and periodic achieving perfect scores and incidental showing slight variability. The time-weighted measure further highlighted task dependence: periodic failed to outperform None in Sudoku but did so in Connect Four, suggesting that periodic updates may interfere more in single-focus intensive tasks. Perceived disruption also diverged, with both tasks rating periodic as more disruptive, but in Sudoku, with a clear advantage for incidental. Workload and primary task performance remained stable across modes. These patterns reveal how task demands can modulate the impact of different visualization strategies.

Overall, IVs provided an effective balance between informativeness and non-disruption. Comparable accuracy and workload to ambient visualizations and lower disruption than periodic indicate that transient, context-aware cues can enhance comprehension while minimizing interference. This aligns with design principles of “information when relevant,” promoting efficient cognitive integration in AR.  
The selected visualization idioms proved effective for brief, incidental viewing due to their perceptually accurate encodings and task relevance \cite{Cleveland:1984, Willett:2017, Quadri:2022}. Participants successfully interpreted visual data even under short exposures, supporting prior work on glanceable visualizations in immersive contexts \cite{Moreira:2020, Moreira:2023a, Moreira:2023b, Moreira:2024}.  

Importantly, these results indicate that augmented reality is a well-suited medium for incidental visualizations. AR provides spatially grounded context for information presentation, while the incidental and transient nature of IVs mitigates common AR challenges such as cognitive overload and visual occlusion. Across both analytical (Sudoku) and interactive (Connect Four) tasks, IVs remained robust, maintaining usability and cognitive efficiency. Together, these findings suggest that AR-based incidental visualizations can deliver high informational value while preserving user focus, achieving a favorable balance between utility and non-intrusiveness.

\begin{comment}
\begin{table}[t]
\centering
\renewcommand{\arraystretch}{1.3}
\setlength{\tabcolsep}{6pt}
\caption{Summary of hypotheses and outcomes across all measures.}
\label{tab:hyp_summary}
\begin{tabular}{|p{1.5cm}|p{.61cm}|p{2.8cm}|p{1.8cm}|}
\hline
\textbf{Metric} & \textbf{Hyp.} & \textbf{Expectation} & \textcolor{darkgray}{\textbf{Outcome}} \\[2pt] \hline
% if needed yellow for partial: yellow!30
\multirow{3}{*}{Accuracy} 
    & \textbf{H1} & Visualizations $>$ None & \cellcolor{green!20}\textbf{Supported} \\
    & \textbf{H2} & incidental $>$ periodic & \cellcolor{red!25}\makecell[l]{\textbf{Not }\\ \textbf{Supported}} \\
    & \textbf{H3} & incidental $\approx$ ambient & \cellcolor{green!20}\textbf{Supported} \\[4pt] \hline

\multirow{2}{*}{Disruption} 
    & \textbf{H4} & None $<$ All & \cellcolor{green!20}\textbf{Supported} \\ 
    & \textbf{H5} & incidental $<$ periodic, ambient & \cellcolor{red!25}\makecell[l]{\textbf{Not }\\ \textbf{Supported}} \\[4pt] \hline

\multirow{2}{*}{Performance} 
    & \textbf{H6} & Visualizations $\approx$ None & \cellcolor{green!20}\textbf{Supported} \\
    & \textbf{H7} & incidental $>$ Others & \cellcolor{red!25}\makecell[l]{\textbf{Not }\\ \textbf{Supported}} \\[2pt] \hline
\end{tabular}
\vspace{0.3em}
\hfill
\parbox{0.99\linewidth}{
    {\footnotesize \textit{Note.} Hyp. = Hypothesis.}
}
\end{table}
\end{comment}

\subsection{Limitations and Future Work}
While the sample size limited parametric analysis, 30 participants provided meaningful insights for AR research. Future studies should include larger, more diverse samples, explore long-term use, and assess IVs in other realistic scenarios. Eye-tracking would further clarify attentional effects, gaze behavior, and distraction in response to transient visualizations \cite{Piening:2021, Davari:2022}.  
Investigating alternative visual encodings (e.g., radial or line charts), more complex or collaborative tasks, and covariates such as perceptual speed \cite{Liu:2020, Quadri:2022} would refine understanding of user variability and cognitive efficiency.  
Ultimately, IVs show promise as contextually adaptive visualizations that minimize occlusion and cognitive load in dynamic AR environments. By combining adaptivity, relevance, and unobtrusiveness, they support informative yet fluid user experiences in immersive analytics.

%% file: sections/conclusions.tex
\section{Conclusion}
\label{chap:conclusion}

In a world increasingly defined by the coexistence of digital and physical realities, information is no longer confined to screens, flowing through our environments, shaping how we act, decide, and understand. Yet, as this increasing presence expands, so does the challenge to deliver data seamlessly, without overloading attention or interrupting natural interactions.

This question lies at the heart of our work, exploring how IVs, as brief, contextually triggered visualizations that appear during ongoing real-world activities, can help bridge this gap. Specifically, it aimed to empirically test and validate the applicability of IVs in AR, focusing on their ability to convey information efficiently while maintaining the user’s primary focus through a controlled study comparing four visualization modes (ambient, incidental, periodic, and none) across two AR tasks. The work examined their effects on accuracy, disruption, workload, and performance.

Results indicated that visualizations can convey data effectively without increasing cognitive load or disrupting task focus. Both Ambient and Incidental modes maintained high task accuracy and low disruption, outperforming the time-based Periodic mode. Across both single-user and collaborative contexts, participants maintained high task accuracy while interpreting incidental information. These findings supported the viability of IVs for context-aware, low-attention information delivery, showing that users can integrate secondary visual cues without compromising their performance or attention allocation. While ambient visualizations achieved the highest accuracy and the lowest perceived disruption, they also carry inherent limitations, most notably occlusion of real-world elements, a persistent challenge in AR design \cite{Davari:2020}. However, the incidental mode matched ambient performance while minimizing visual clutter and occlusion, confirming that transient, context-aware cues can deliver information efficiently and unobtrusively. 

These findings demonstrated that visualizations in AR need not dominate attention to be effective. When designed around context, timing, and perceptual simplicity, they can integrate naturally into the user's workflows, enhancing understanding while preserving focus and awareness. In doing so, our work further positions IVs as a promising paradigm for delivering contextually relevant, glanceable information that harmonizes, rather than competes against, the dynamics of human attention and expanding its application range to AR environments.

%% file: egbibsample.bib
@ARTICLE{Bressa:2021,
    author={Bressa, Nathalie and Korsgaard, Henrik and Tabard, Aurélien and Houben, Steven and Vermeulen, Jo},
    journal={IEEE Transactions on Visualization and Computer Graphics}, 
    title={What's the Situation with Situated Visualization? A Survey and Perspectives on Situatedness}, 
    year={2022},
    volume={28},
    number={1},
    pages={107-117},
    doi={10.1109/TVCG.2021.3114835}
}

@ARTICLE{Kraus:2021,
    author={Kraus, Matthias and Klein, Karsten and Fuchs, Johannes and Keim, Daniel A. and Schreiber, Falk and Sedlmair, Michael},
    journal={IEEE Computer Graphics and Applications}, 
    title={The Value of Immersive Visualization}, 
    year={2021},
    volume={41},
    number={4},
    pages={125-132},
    doi={10.1109/MCG.2021.3075258}
}

@article{Martins:2022,
    author       = {Nuno Cid Martins and
                  Bernardo Marques and
                  Jo{\~{a}}o Alves and
                  Tiago Ara{\'{u}}jo and
                  Paulo Dias and
                  Beatriz Sousa Santos},
    title        = {Augmented reality situated visualization in decision-making},
    journal      = {Multim. Tools Appl.},
    volume       = {81},
    number       = {11},
    pages        = {14749--14772},
    year         = {2022},
    url          = {https://doi.org/10.1007/s11042-021-10971-4},
    doi          = {10.1007/S11042-021-10971-4},
    bibsource    = {dblp computer science bibliography, https://dblp.org}
}

@inproceedings{Lu:2021.1_EvaluatingPotential,
      author       = {Feiyu Lu and
                      Doug A. Bowman},
      title        = {Evaluating the Potential of Glanceable {AR} Interfaces for Authentic Everyday Uses},
      booktitle    = {{IEEE} Virtual Reality and 3D User Interfaces, {VR} 2021, Lisbon,
                      Portugal, March 27 - April 1, 2021},
      pages        = {768--777},
      publisher    = {{IEEE}},
      year         = {2021},
      url          = {https://doi.org/10.1109/VR50410.2021.00104},
      doi          = {10.1109/VR50410.2021.00104},
      bibsource    = {dblp computer science bibliography, https://dblp.org}
}

@inproceedings{Lu:2021.2_Exploration_Techniques,
    author       = {Feiyu Lu and
                  Shakiba Davari and
                  Doug A. Bowman},
    editor       = {Francisco R. Ortega and
                  Robert J. Teather and
                  Gerd Bruder and
                  Thammathip Piumsomboon and
                  Benjamin Weyers and
                  Anil Ufuk Batmaz and
                  Kyle Johnsen and
                  Christoph W. Borst},
    title        = {Exploration of Techniques for Rapid Activation of Glanceable Information in Head-Worn Augmented Reality},
    booktitle    = {{SUI} '21: Symposium on Spatial User Interaction, Virtual Event, USA,
                  November 9-10, 2021},
    pages        = {14:1--14:11},
    publisher    = {{ACM}},
    year         = {2021},
    url          = {https://doi.org/10.1145/3485279.3485286},
    doi          = {10.1145/3485279.3485286},
    bibsource    = {dblp computer science bibliography, https://dblp.org}
}

@article{Jeffri:2021,
    author = {Jeffri, Nor Farzana and Rambli, Dayang},
    year = {2021},
    month = {03},
    pages = {e06277},
    title = {A review of augmented reality systems and their effects on mental workload and task performance},
    volume = {7},
    journal = {Heliyon},
    doi = {10.1016/j.heliyon.2021.e06277}
}

@inproceedings{Marques:2019,
  title = {Situated Visualization in The Decision Process Through Augmented Reality},
  url = {http://dx.doi.org/10.1109/IV.2019.00012},
  DOI = {10.1109/iv.2019.00012},
  booktitle = {2019 23rd International Conference Information Visualisation (IV)},
  publisher = {IEEE},
  author = {Marques,  Bernardo and Santos,  Beatriz Sousa and Araujo,  Tiago and Martins,  Nuno Cid and Alves,  Joao Bernardo and Dias,  Paulo},
  year = {2019},
  month = jul,
  pages = {13–18}
}

@article{Lee:2024,
    author       = {Benjamin Lee and
                  Michael Sedlmair and
                  Dieter Schmalstieg},
    title        = {Design Patterns for Situated Visualization in Augmented Reality},
    journal      = {{IEEE} Trans. Vis. Comput. Graph.},
    volume       = {30},
    number       = {1},
    pages        = {1324--1335},
    year         = {2024},
    url          = {https://doi.org/10.1109/TVCG.2023.3327398},
    doi          = {10.1109/TVCG.2023.3327398},
    bibsource    = {dblp computer science bibliography, https://dblp.org}
}

@inproceedings{Ens:2021,
    author = {Ens, Barrett and Bach, Benjamin and Cordeil, Maxime and Engelke, Ulrich and Serrano, Marcos and Willett, Wesley and Prouzeau, Arnaud and Anthes, Christoph and B\"{u}schel, Wolfgang and Dunne, Cody and Dwyer, Tim and Grubert, Jens and Haga, Jason H. and Kirshenbaum, Nurit and Kobayashi, Dylan and Lin, Tica and Olaosebikan, Monsurat and Pointecker, Fabian and Saffo, David and Saquib, Nazmus and Schmalstieg, Dieter and Szafir, Danielle Albers and Whitlock, Matt and Yang, Yalong},
    title = {Grand Challenges in Immersive Analytics},
    year = {2021},
    isbn = {9781450380966},
    publisher = {Association for Computing Machinery},
    address = {New York, NY, USA},
    url = {https://doi.org/10.1145/3411764.3446866},
    doi = {10.1145/3411764.3446866},
    booktitle = {Proceedings of the 2021 CHI Conference on Human Factors in Computing Systems},
    articleno = {459},
    numpages = {17},
    location = {Yokohama, Japan},
    series = {CHI '21}
}

@inproceedings{Moreira:2020,
    author       = {Jo{\~{a}}o Moreira and
                  Daniel Mendes and
                  Daniel Gon{\c{c}}alves},
    editor       = {Genny Tortora and
                  Giuliana Vitiello and
                  Marco Winckler},
    title        = {Incidental Visualizations: Pre-Attentive Primitive Visual Tasks},
    booktitle    = {{AVI} '20: International Conference on Advanced Visual Interfaces,
                  Island of Ischia, Italy, September 28 - October 2, 2020},
    pages        = {25:1--25:9},
    publisher    = {{ACM}},
    year         = {2020},
    url          = {https://doi.org/10.1145/3399715.3399841},
    doi          = {10.1145/3399715.3399841},
    bibsource    = {dblp computer science bibliography, https://dblp.org}
}

@article{Moreira:2024,
    title = {Incidental visualizations: How complexity factors influence task performance},
    journal = {Visual Informatics},
    volume = {8},
    number = {4},
    pages = {85-96},
    year = {2024},
    issn = {2468-502X},
    doi = {https://doi.org/10.1016/j.visinf.2024.10.005},
    url = {https://www.sciencedirect.com/science/article/pii/S2468502X24000652},
    author = {João Moreira and Daniel Mendes and Daniel Gonçalves}
}

@inproceedings{Blascheck:2021.1,
    author       = {Tanja Blascheck and
                  Petra Isenberg},
    editor       = {Christophe Hurter and
                  Helen C. Purchase and
                  Jos{\'{e}} Braz and
                  Kadi Bouatouch},
    title        = {A Replication Study on Glanceable Visualizations: Comparing Different Stimulus Sizes on a Laptop Computer},
    booktitle    = {Proceedings of the 16th International Joint Conference on Computer
                  Vision, Imaging and Computer Graphics Theory and Applications, {VISIGRAPP}
                  2021, Volume 3: IVAPP, Online Streaming, February 8-10, 2021},
    pages        = {133--143},
    publisher    = {{SCITEPRESS}},
    year         = {2021},
    url          = {https://doi.org/10.5220/0010328501330143},
    doi          = {10.5220/0010328501330143},
    bibsource    = {dblp computer science bibliography, https://dblp.org}
}

@incollection{Blascheck:2021.2_Characterizing_GlanceableVis,
  TITLE = {{Characterizing Glanceable Visualizations: From Perception to Behavior Change}},
  AUTHOR = {Blascheck, Tanja and Bentley, Frank and Choe, Eun Kyoung and Horak, Tom and Isenberg, Petra},
  URL = {https://inria.hal.science/hal-03524091},
  BOOKTITLE = {{Mobile Data Visualization}},
  PUBLISHER = {{Chapman and Hall/CRC}},
  PAGES = {151-176},
  YEAR = {2021},
  MONTH = Nov,
  DOI = {10.1201/9781003090823-5},
  HAL_ID = {hal-03524091},
  HAL_VERSION = {v1},
}

@inproceedings{Piening:2021,
    author       = {Robin Piening and
                  Ken Pfeuffer and
                  Augusto Esteves and
                  Tim Mittermeier and
                  Sarah Prange and
                  Philippe Schr{\"{o}}der and
                  Florian Alt},
    editor       = {Carmelo Ardito and
                  Rosa Lanzilotti and
                  Alessio Malizia and
                  Helen Petrie and
                  Antonio Piccinno and
                  Giuseppe Desolda and
                  Kori Inkpen},
    title        = {Looking for Info: Evaluation of Gaze Based Information Retrieval in Augmented Reality},
    booktitle    = {Human-Computer Interaction - {INTERACT} 2021 - 18th {IFIP} {TC} 13
                  International Conference, Bari, Italy, August 30 - September 3, 2021,
                  Proceedings, Part {I}},
    series       = {Lecture Notes in Computer Science},
    volume       = {12932},
    pages        = {544--565},
    publisher    = {Springer},
    year         = {2021},
    url          = {https://doi.org/10.1007/978-3-030-85623-6\_32},
    doi          = {10.1007/978-3-030-85623-6\_32},
    bibsource    = {dblp computer science bibliography, https://dblp.org}
}

@article{Cleveland:1984,
    Author = {William S. Cleveland and Robert McGill},
    title = {Graphical Perception: Theory, Experimentation, and Application to the Development of Graphical Methods},
    journal = {Journal of the American Statistical Association},
    volume = {79},
    number = {387},
    pages = {531--554},
    year = {1984},
    publisher = {ASA Website},
    doi = {10.1080/01621459.1984.10478080},
    URL = {    https://www.tandfonline.com/doi/abs/10.1080/01621459.1984.10478080
    },
    eprint = { https://www.tandfonline.com/doi/pdf/10.1080/01621459.1984.10478080
    }
}

@article{Carswell:1987,
    author = {C. Melody Carswell and Christopher D. Wickens},
    title = {Information integration and the object display An interaction of task demands and display superiority},
    journal = {Ergonomics},
    volume = {30},
    number = {3},
    pages = {511--527},
    year = {1987},
    publisher = {Taylor \& Francis},
    doi = {10.1080/00140138708969741}, 
    URL = { 
        
            https://doi.org/10.1080/00140138708969741
    },
    eprint = {     
            https://doi.org/10.1080/00140138708969741
    }
}

@article{Moreira:2023a,
    author = {João Moreira and Daniel Mendes and Daniel Gonçalves},
    title ={Impact of incidental visualizations on primary tasks},
    journal = {Information Visualization},
    volume = {22},
    number = {4},
    pages = {307-322},
    year = {2023},
    doi = {10.1177/14738716231180892},
    URL = { 
            https://doi.org/10.1177/14738716231180892
    },
    eprint = {    
            https://doi.org/10.1177/14738716231180892
    }
}

@article{Moreira:2023b,
    author = {João Moreira and Daniel Mendes and Daniel Gonçalves},
    title ={Incidental graphical perception: How marks and display time influence accuracy},
    journal = {Information Visualization},
    volume = {23},
    number = {1},
    pages = {3-20},
    year = {2024},
    doi = {10.1177/14738716231189218},
    URL = { 
            https://doi.org/10.1177/14738716231189218
     },
    eprint = { 
            https://doi.org/10.1177/14738716231189218
    }
}

@INPROCEEDINGS{Moreira:2024b_IVvsAMB,
    author={Moreira, João and Pinto, Diogo and Mendes, Daniel and Gonçalves, Daniel},
    booktitle={2024 International Conference on Graphics and Interaction (ICGI)}, 
    title={Incidental Versus Ambient Visualizations: Comparing Cognitive and Mechanical Tasks}, 
    year={2024},
    volume={},
    number={},
    pages={1-8},
    doi={10.1109/ICGI64003.2024.10923794}
}

@incollection{Hart:1988,
    title = {Development of NASA-TLX (Task Load Index): Results of Empirical and Theoretical Research},
    editor = {Peter A. Hancock and Najmedin Meshkati},
    series = {Advances in Psychology},
    publisher = {North-Holland},
    volume = {52},
    pages = {139-183},
    year = {1988},
    booktitle = {Human Mental Workload},
    issn = {0166-4115},
    doi = {https://doi.org/10.1016/S0166-4115(08)62386-9},
    url = {https://www.sciencedirect.com/science/article/pii/S0166411508623869},
    author = {Sandra G. Hart and Lowell E. Staveland},
}

@article{KleinSedlmairSchreiber:2022,
    url = {https://doi.org/10.1515/itit-2022-0037},
    title = {Immersive analytics: An overview},
    author = {Karsten Klein and Michael Sedlmair and Falk Schreiber},
    pages = {155--168},
    volume = {64},
    number = {4-5},
    journal = {it - Information Technology},
    doi = {doi:10.1515/itit-2022-0037},
    year = {2022},
    lastchecked = {2024-12-28}
}

@INPROCEEDINGS{Davari:2022,
  author={Davari, Shakiba and Lu, Feiyu and Bowman, Doug A.},
  booktitle={2022 IEEE Conference on Virtual Reality and 3D User Interfaces (VR)}, 
  title={Validating the Benefits of Glanceable and Context-Aware Augmented Reality for Everyday Information Access Tasks}, 
  year={2022},
  volume={},
  number={},
  pages={436-444},
  doi={10.1109/VR51125.2022.00063}}

@inproceedings{Daskalogrigorakis:2021,
    author       = {Grigoris Daskalogrigorakis and
              Ann McNamara and
              Katerina Mania},
    title        = {Holo-Box: Level-of-Detail Glanceable Interfaces for Augmented Reality},
    booktitle    = {{SIGGRAPH} 2021: Special Interest Group on Computer Graphics and Interactive
              Techniques Conference, Posters, Virtual Event, USA, August 9-13, 2021},
    pages        = {13:1--13:2},
    publisher    = {{ACM}},
    year         = {2021},
    url          = {https://doi.org/10.1145/3450618.3469175},
    doi          = {10.1145/3450618.3469175},
    bibsource    = {dblp computer science bibliography, https://dblp.org}
}

@inproceedings{Iquiapaza:2023,
    author       = {Yhonatan Iquiapaza and
                  Jorge A. Wagner Filho and
                  Luciana P. Nedel},
    title        = {DeAR: Combining Desktop and Augmented Reality for Visual Data Analysis},
    booktitle    = {Proceedings of the 25th Symposium on Virtual and Augmented Reality,
                  {SVR} 2023, Rio Grande, Brazil, November 6-9, 2023},
    pages        = {233--237},
    publisher    = {{ACM}},
    year         = {2023},
    url          = {https://doi.org/10.1145/3625008.3625021},
    doi          = {10.1145/3625008.3625021},
    bibsource    = {dblp computer science bibliography, https://dblp.org}
}

@article{Pohl:2024,
    author       = {Henning Pohl},
    title        = {Body-Based Augmented Reality Feedback During Conversations},
    journal      = {Proc. {ACM} Hum. Comput. Interact.},
    volume       = {8},
    number       = {{MHCI}},
    pages        = {1--22},
    year         = {2024},
    url          = {https://doi.org/10.1145/3676491},
    doi          = {10.1145/3676491},
    bibsource    = {dblp computer science bibliography, https://dblp.org}
}

@article{Willett:2017,
    author       = {Wesley Willett and
                  Yvonne Jansen and
                  Pierre Dragicevic},
    title        = {Embedded Data Representations},
    journal      = {{IEEE} Trans. Vis. Comput. Graph.},
    volume       = {23},
    number       = {1},
    pages        = {461--470},
    year         = {2017},
    url          = {https://doi.org/10.1109/TVCG.2016.2598608},
    doi          = {10.1109/TVCG.2016.2598608},
    bibsource    = {dblp computer science bibliography, https://dblp.org}
}

@inproceedings{Davari:2020,
  author       = {Shakiba Davari and
                  Feiyu Lu and
                  Doug A. Bowman},
  title        = {Occlusion Management Techniques for Everyday Glanceable {AR} Interfaces},
  booktitle    = {2020 {IEEE} Conference on Virtual Reality and 3D User Interfaces Abstracts
                  and Workshops, {VR} Workshops, Atlanta, GA, USA, March 22-26, 2020},
  pages        = {324--330},
  publisher    = {{IEEE}},
  year         = {2020},
  url          = {https://doi.org/10.1109/VRW50115.2020.00072},
  doi          = {10.1109/VRW50115.2020.00072},
  bibsource    = {dblp computer science bibliography, https://dblp.org}
}

@article{Liu:2020,
    author = {Liu, Zhengliang and Crouser, R. Jordan and Ottley, Alvitta},
    title = {Survey on Individual Differences in Visualization},
    journal = {Computer Graphics Forum},
    volume = {39},
    number = {3},
    pages = {693-712},
    doi = {https://doi.org/10.1111/cgf.14033},
    url = {https://onlinelibrary.wiley.com/doi/abs/10.1111/cgf.14033},
    eprint = {https://onlinelibrary.wiley.com/doi/pdf/10.1111/cgf.14033},
    year = {2020}
}

@ARTICLE{Quadri:2022,
    author={Quadri, Ghulam Jilani and Rosen, Paul},
    journal={IEEE Transactions on Visualization and Computer Graphics}, 
    title={A Survey of Perception-Based Visualization Studies by Task}, 
    year={2022},
    volume={28},
    number={12},
    pages={5026-5048},
    doi={10.1109/TVCG.2021.3098240}
}

@inproceedings{Imamov:2020,
    author       = {Samat Imamov and
                  Daniel Monzel and
                  Wallace Santos Lages},
    title        = {Where to display? How Interface Position Affects Comfort and Task
                  Switching Time on Glanceable Interfaces},
    booktitle    = {{IEEE} Conference on Virtual Reality and 3D User Interfaces, {VR}
                  2010, Atlanta, GA, USA, March 22-26, 2020},
    pages        = {851--858},
    publisher    = {{IEEE}},
    year         = {2020},
    url          = {https://doi.org/10.1109/VR46266.2020.1581435674325},
    doi          = {10.1109/VR46266.2020.1581435674325},
    bibsource    = {dblp computer science bibliography, https://dblp.org}
}

@article{Vanderplas:2020,
    title = "Testing statistical charts: what makes a good graph?",
    author = "Susan Vanderplas and Dianne Cook and Heike Hofmann",
    year = "2020",
    month = mar,
    day = "7",
    doi = "10.1146/annurev-statistics-031219-041252",
    language = "English",
    volume = "7",
    pages = "61--88",
    journal = "Annual Review of Statistics and Its Application",
    issn = "2326-8298",
    publisher = "Annual Reviews",

}

@inproceedings{Guarese:2020,
    author = {Guarese, Renan and Becker, Jo\~{a}o and Fensterseifer, Henrique and Walter, Marcelo and Freitas, Carla and Nedel, Luciana and Maciel, Anderson},
    title = {Augmented Situated Visualization for Spatial and Context-Aware Decision-Making},
    year = {2020},
    isbn = {9781450375351},
    publisher = {Association for Computing Machinery},
    address = {New York, NY, USA},
    url = {https://doi.org/10.1145/3399715.3399838},
    doi = {10.1145/3399715.3399838},
    booktitle = {Proceedings of the 2020 International Conference on Advanced Visual Interfaces},
    articleno = {48},
    numpages = {5},
    location = {Salerno, Italy},
    series = {AVI '20}
}

@inbook{Weiser_Brown:1996,
  title = {Das kommende Zeitalter der Calm Technology},
  ISBN = {9783839430460},
  ISSN = {2702-8860},
  url = {http://dx.doi.org/10.14361/9783839430460-001},
  DOI = {10.14361/9783839430460-001},
  booktitle = {Internet der Dinge},
  publisher = {transcript Verlag},
  author = {Weiser,  Mark and Brown,  John Seely},
  year = {2015},
  month = sep,
  pages = {59–72}
}

@article{Weiser:1993,
    author = {Weiser, Mark},
    title = {Some computer science issues in ubiquitous computing},
    year = {1993},
    issue_date = {July 1993},
    publisher = {Association for Computing Machinery},
    address = {New York, NY, USA},
    volume = {36},
    number = {7},
    issn = {0001-0782},
    url = {https://doi.org/10.1145/159544.159617},
    doi = {10.1145/159544.159617},
    journal = {Commun. ACM},
    month = jul,
    pages = {75–84},
    numpages = {10}
}

@inproceedings{Ishii:1998,
    author = {Wisneski, Craig and Ishii, Hiroshi and Dahley, Andrew and Gorbet, Matthew G. and Brave, Scott and Ullmer, Brygg and Yarin, Paul},
    title = {Ambient Displays: Turning Architectural Space into an Interface between People and Digital Information},
    year = {1998},
    isbn = {3540642374},
    publisher = {Springer-Verlag},
    address = {Berlin, Heidelberg},
    booktitle = {Proceedings of the First International Workshop on Cooperative Buildings, Integrating Information, Organization, and Architecture},
    pages = {22–32},
    numpages = {11},
    series = {CoBuild '98}
}

@misc{pygame,
    author       = {{Pygame Community}},
    title        = {Pygame -- Python Game Development},
    year         = {2000--2025},
    howpublished = {\url{https://www.pygame.org/}},
    note         = {Version 2.6.1, accessed 24 August 2025},
    license      = {LGPL}
}

@misc{JustA3DObject:2023,
    author       = {JustA3DObject},
    title        = {Connect Four Game},
    year         = {2023},
    howpublished = {\url{https://github.com/gdsc-ipsacademy/Connect-Four-Game/blob/main/src/game.py}},
    note         = {MIT License}
}

@misc{LaerdStatistics2015,
  author       = {{Laerd Statistics}},
  title        = {Friedman Test using SPSS Statistics},
  year         = {2015},
  howpublished = {\url{https://statistics.laerd.com/statistical-guides/friedman-test-using-spss-statistics.php}},
  note         = {Statistical tutorials and software guides. Retrieved from https://statistics.laerd.com/}
}
